\documentclass{aa}  

\usepackage{graphicx}
\usepackage{placeins}
\usepackage{afterpage}
\usepackage{natbib}
\usepackage{enumitem}
\usepackage{siunitx}
\usepackage{svg}
\usepackage{tabularx}
\usepackage{multirow}
\usepackage{url}
\usepackage{float}
\usepackage{amsmath}\usepackage{bm}
\usepackage{adjustbox}
\usepackage{verbatim} 
\usepackage{bbm}
\usepackage[varg]{txfonts}
\usepackage[hidelinks]{hyperref}
\usepackage{xcolor}

\hypersetup{
    colorlinks,
    linkcolor={black!50!black},
    citecolor={blue!50!black},
    urlcolor={blue!80!black}
}

\makeatletter
\renewcommand*\aa@manuscriptname{%
  manuscript no. aa59974-26%
  \hspace{\stretch{1}}%
  \copyright ESO \the\year
}
\makeatother

\begin{document}

   \title{Accretion across scales: streamers, surface-layer transport, and rapid replenishment in young protoplanetary discs }
    
\titlerunning{Accretion across scales}
   \authorrunning{Christian Granzow Holm et al.}

   \author{C. Granzow Holm
          \inst{1}
          \and
          M. Lambrechts\inst{1}
          \and
          M. Kuffmeier\inst{2}
          \and
          A. Johansen\inst{1}
          \and
          T. Haugbølle\inst{2}
          \and
          Å. Nordlund\inst{2}
          }

   \institute{Centre for Star and Planet Formation, Globe Institute, University of Copenhagen,
              Øster Voldgade 5–7, 1350 Copenhagen K, Denmark\\
              \email{christian.holm@sund.ku.dk}
         \and
             Niels Bohr Institute, University of Copenhagen, 
             Jagtvej 155A, 2200 Copenhagen N, Denmark\\}

   \date{Received 20 March 2026 / Accepted 3 August 2026}

 
  \abstract{
Protoplanetary discs evolve around newly-formed stars through an interplay of infall from surrounding turbulent cloud material, accretion towards the young star, and outflow driven mass-loss.
It has been challenging to determine if discs are fed predominantly through infall along the disc midplane, or along the poles, and if accretion occurs in a steady or burst-like fashion.
Here, we present a suite of 3D ideal magnetohydrodynamical simulations of protoplanetary disc formation and evolution in a dynamic, large-scale molecular cloud environment using the adaptive mesh refinement framework \textsc{dispatch}.
We focus on nine stellar systems, where we resolve discs down to a scale of $0.8$\,au.
Across the sample, stellar accretion proceeds at rates of $\sim$10$^{-5}$\,M$_\odot$\,yr$^{-1}$ over 10$^{5}$ yr, with significant variability. 
Discs grow to 100\,au scales and remain gravitationally stable in time, with disc-to-star mass ratios below 10\,\%.
Transient high-density streamers, with $10$\,kyr infall times, can drive anisotropic mass delivery at rates comparable to the background accretion flow. 
Their interaction with discs typically results in a temporary reduction of the disc size by half, and disc mass by 40\,\%.
During later quiescent disc evolution stages ($t\gtrsim$50\,kyr), accretion predominantly occurs through the midplane and disc surface layers. This is associated with the development of a toroidal magnetic field morphology, which includes field reversals across both disc surfaces.
In this way, the full disc mass reservoir is replenished on 10\,kyr-timescales.
%
These findings support that the outer parts of very young discs, when well-ionised and close to the ideal MHD regime, are not yet conducive to planet formation, due to high replenishment rates, strong turbulence, and disruptive streamer infall events.
}


   \maketitle
%
\section{Introduction}

\begin{figure*}
    \centering
    \resizebox{\hsize}{!}{\includegraphics{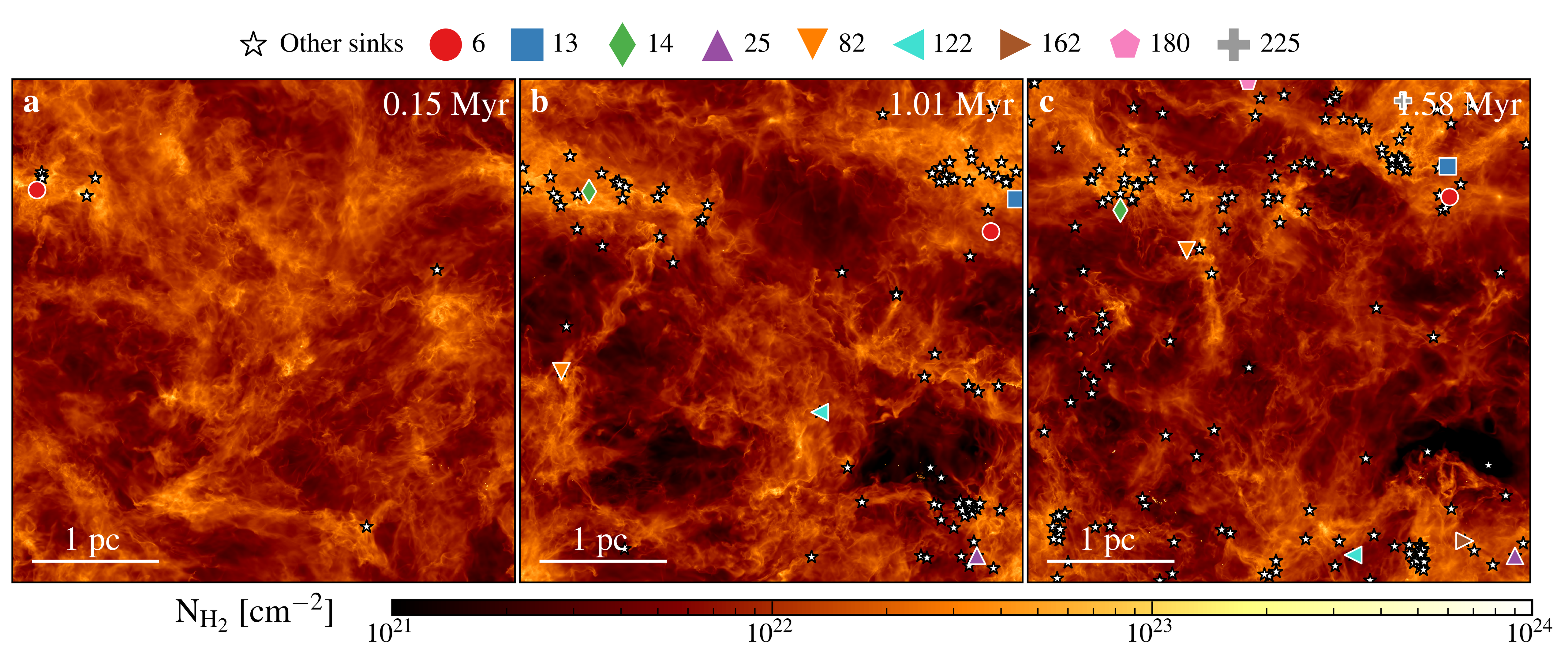}}
    \caption{Time evolution of the column density of the full simulation domain showing the creation of sink particles, including the 9 systems analysed in this paper (see symbols above the plots). 
    Panel a: The formation of the first generation of sinks, including sink 6 (red circle), at a time $t_{\rm Global}=0.15$\,Myr 
    of the global parental run (Sec.\,\ref{sec:methods_parental_run}). 
    Panel b: Snapshot in time subsequent to the formation of sink 122, at $t_{\rm Global}=1.01$\,Myr. 
    Panel c: Snapshot at the creation time of sink 225, $t_{\rm Global}=1.58$\,Myr.
    }
    \label{fig:panel}
\end{figure*}

The formation of a star and its surrounding protoplanetary disc are intertwined processes within the turbulent regions of molecular clouds \citep{padoan_2002_starformation, Mac_Low_2004_starformation}.
As gravitationally unstable dense filaments nucleate to form stars, continued infalling material leads to the formation of protoplanetary discs that are the birthplace for planets. 
These young protoplanetary discs are observed to form with a wide range of radii, 
ranging from $10$\,au to $1000$\,au based on their dust emission \citep{tobin_2020_discdiversity}, and 
with masses between $0.01$\,M$_\odot$ and 
$0.1$\,M$_\odot$, assuming solar dust-to-gas ratios \citep{tychoniec_2020_dustmass}.
This diversity is inferred to be inherited from the turbulent molecular clouds in which they form \citep{kuffmeier_2017, Bate_2018}.


Magnetic fields play a vital role in shaping the formation and evolution of the protoplanetary disc.
During the earliest stages of star formation, magnetic braking due to inwards-dragged magnetic fields was originally proposed to hinder disc formation \citep{basu_1994, allen_2003, mellon_2008_magneticbraking}, unless diffused by ambipolar diffusion \citep{masson_2016_discsizes,mayer_2025_largescale_niMHD_structures} or, more slowly, through turbulence-induced misalignment between magnetic fields and the rotational axis of the disc \citep{hennebelle_2009_disc_misalignment, seifried_2012_turbulence_and_Bfield, wurster_2019_ni_i_disc_sizes}.
On the scales of discs, outflows and magnetic stresses transport angular momentum vertically, 
driving inwards mass transport, under ideal \citep{konigl_pudritz_2000, suzuki_2014_surface_accretion,zhu_stone_2018} and non-ideal 
magnetohydrodynamic (MHD) 
conditions \citep{bai_2013,lesur_2014_thanatology,gressel_2015_MHDdiskwinds}.
In the midplane, 
MHD turbulence, if present, acts as an effective disc viscosity, transferring material radially inwards \citep{shakura_Sunyaev, balbus_hawley_1998}. 
Finally, close to the host star, magnetically-collimated jets inject energy and momentum back into the surrounding medium \citep{machida_2007_jetlunchpoint,banerjee_2007_jetsenv}.

This process of disc formation and early evolution does not occur in isolation, but proceeds through continuous filamentary infall from the surrounding environment \citep{seifried_2013_turbulence_and_Bfield,padoan_2014_filamentary, kuffmeier_2017, lebreuilly_2024a, yang_2025}.
Indeed, discs around young stars are seen to be connected to long filamentary gas overdensities, dubbed streamers, on scales of hundreds to thousands of au 
\citep{yen_2019, alves_2020, pineda_2020_huge_class0_streamer_mass, flores_2023_obs_streamers}.
Such sub-parsec filamentary flows deliver mass and angular momentum \citep{banerjee_2006_cloud2disc}, influencing the disc size and orientation \citep{Kuffmeier_2021_misaligneddisk}.

In this work, we aim to model disc formation in typical Perseus-like star-forming environments with mean cloud column densities of $3\times10^{21}$\,cm$^{-2}$ \citep{hatchell_2005_perseus_starformation}, as opposed to denser star formation environments of $ 5.2\times10^{22}$\,cm$^{-2}$ \citep{Bate_2018} or higher \citep{lebreuilly_2024a}, that are observed in the centre of molecular cloud filaments \citep{andre_2014_nonspherical_cores}.
Additionally, we aim to cover the main accretion phase during Class 0, corresponding to 10$^5$\,yr of evolution. Therefore, we need parsec-scale simulation domains to capture accretion regions that can exceed $0.1$\,pc \citep{kuffmeier_2023}.
For this, we use ideal MHD zoom-in simulations within a (4\,pc)$^3$ simulation domain to explore the initial stage where stars and discs form ($t\lesssim 10^4$\,yr) and the further evolution towards protostellar evolution timescales, $t\approx 10^4 - 10^5$\,yr, where the still-embedded star-disc system continues to accrete gas from the surrounding medium.
During this transition, we seek to characterise the environmental impact on disc structure and the pathways through which gas accretes onto the disc and the protostar.

The paper is structured as follows. 
Section\,\ref{sec:methods} describes the methodology of our zoom-in simulations, which start from a molecular cloud simulation similar to \citet{haugbølle_2018}\,
performed with \textsc{ramses} \citep{teyssier_2001}.
We then select nine systems for closer inspection, similar to the zoom-in methodology of \citet{kuffmeier_2017,kuffmeier_2018}.
These systems are then re-simulated using the adaptive mesh refinement (AMR) code framework \textsc{dispatch} \citep{nordlund_2018}.
The properties of the nine prestellar cores are given in Section\,\ref{sec:resIC}. 
Section\,\ref{sec:discformevol} focuses on the disc formation and evolution of a relatively isolated system, highlighting the fast replenishment of gas through the disc.
We place this result subsequently in the context of the large diversity between the evolution of different systems, with particular focus on massive streamer infall events (Sec.\,\ref{sec:results_disk_diversity}). In Section\,\ref{sec:discussion_previous_work}, we discuss previous work and potential effects from radiative transfer and non-ideal MHD.
We argue that young discs are not yet conducive to planet formation (Sec.\,\ref{sec:discussion_implications_for_planet_formation}).
Finally, our conclusions are presented in Section\,\ref{sec:conclusion}.
The appendices provide an expanded description of our methods (Appendix\,\ref{sec:appendix_methods}) and complementary analysis (Appendix\,\ref{appendix:complementary_analysis}).

\section{Methods and simulation setup}\label{sec:methods}

\begin{figure*}
    \centering
    \resizebox{\hsize}{!}{\includegraphics{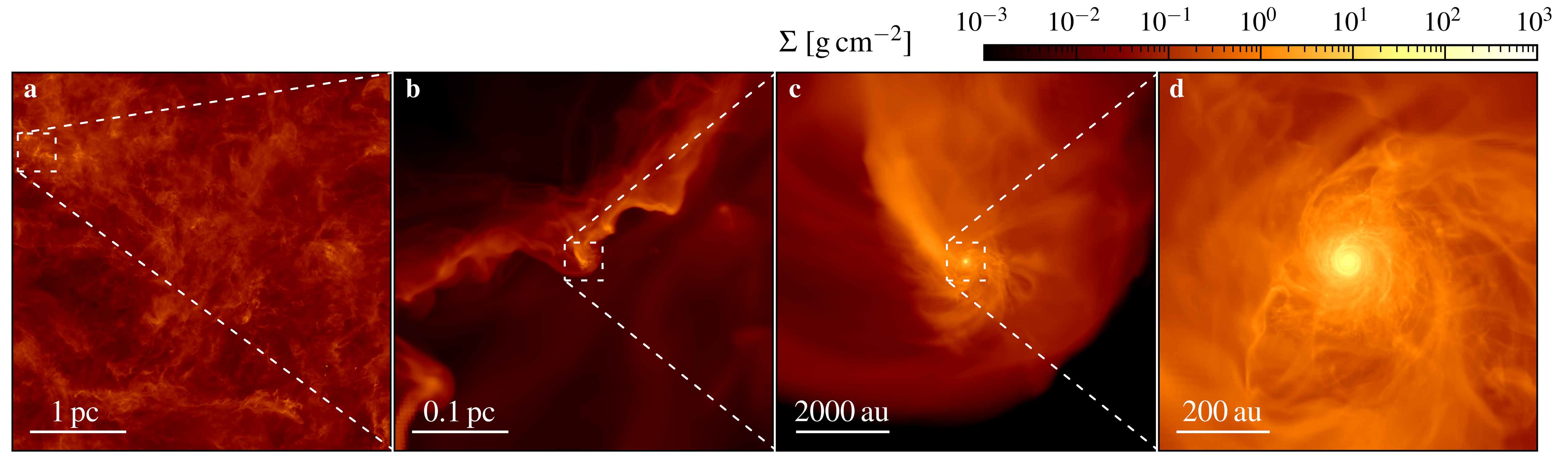}}
    \caption{Integrated column densities ($\Sigma$) at four different scales, zooming in on system \texttt{13} at $t=62$\,kyr.
    The integration depths are: $4$\,pc for panel a, $0.4$\,pc for panel b, $0.04$\,pc for panel c, and $0.004$\,pc for panel d.
    Panel (a) displays the full computational domain with an average column density of $\Sigma = 3.9 \times 10^{-2}$\,g\,cm$^{-2}$.
    Zooming in on \texttt{13}, panel (b) shows how the star and disc form along dense filaments on scales of the Jeans length, $\lambda_{\rm L}=0.12$\,pc.
    On even smaller scales, panel (c) reveals the infalling envelope 4000\,au in size.
    On the outermost right plot, panel d, a clear disc structure is visible with a radius of about $100$\,au.
    }
    \label{fig:zoomin}
\end{figure*}

We select nine newly-formed stars from a large-scale parental (4\,pc)$^3$ star formation simulation described in Section\,\ref{sec:methods_parental_run}.
These systems are then re-simulated for time intervals of about 10$^5$\,yr in the task-based code framework \textsc{dispatch} \citep[][see Sec.\,\ref{sec:dispatch}]{nordlund_2018}. For each system, we preserve the full parent box domain, but transform to the rest-frame of the star and employ a combination of geometric and physical refinement criteria to reach a minimum cell size of approximately $0.8$\,au close to the young star.
In this way, we sample different core properties, physically inherited from a realistic larger environment, as opposed to an approach using systematic parameter variations for isolated systems with idealised initial and boundary conditions.

\subsection{Physical model and numerical methods in \textsc{dispatch}}
\label{sec:dispatch}

We use \textsc{dispatch} to perform simulations in the ideal MHD regime, with a barotropic equation of state,  self-gravity, and use a sub-cell sink particle model representing stars.
The ideal MHD equations are solved with the approximate Riemann solver Harten-Lax-van Leer discontinuities (HLLD) method \citep[][Appendix\,\ref{appendix:MHD_equations}]{miyoshi_2005,Fromang2006}. 
The standard piece-wise barotropic equation of state is described in more detail in Appendix\,\ref{appendix:Eos}.
The sink particle prescription is similar to \citet{haugbølle_2018} and fully described in Appendix\,\ref{appendix:sink_cell_prescription}.
The gravitational potential $\Phi$ includes only the contribution of the gas mass. The multigrid method used for solving the corresponding Poisson equation is described in \citet{Ramsey_2018}. The forces between sink particles are calculated with an $N$-body method, where time interpolation is used to make the method approximately reflexive in time. The gravitational forces between sink particles and the gas are computed by direct
summation and accumulated simultaneously for sinks and cells, resulting in exact conservation in the momentum exchange.
Periodically a Galilean transform is applied to keep the simulation in the rest frame of the sink particle and reduce numerical diffusion from supersonic bulk advection of the disc-star systems.

\textsc{dispatch} segments the simulation domain into patches \citep{nordlund_2018}. 
To optimise cache reuse and load-balancing of the task queue in our simulations, we chose a patch size of 16$^3$ cells. To capture the gravitational collapse and formation of the protostar, while retaining the larger-scale environment, we use a combination of geometric, local Jeans, and vorticity criteria for refinement. A necessary condition for refining a patch is that the distance to the central sink, or the centre of the prestellar core, is less than 144 cells at any level of refinement. In \textsc{dispatch}, refinement criteria are checked locally in each octant of a patch and can result in either no change of refinement, derefinement or refinement, the latter resulting in the creation of between 1 and 8 child patches. Furthermore, refinement or existence of a patch can be mandated to maintain a ``graded'' refinement, with no more than one level difference in resolution between neighbouring patches. In each patch, the normalised number of cells per Jeans length and vorticity scale are computed as
\begin{align}
R_J &= \frac{L_{\rm J, min}}{L_J} \\
R_\omega &= \frac{\Delta s\,|\vec{\nabla}\times\vec{v}|}{\mathrm{\omega_{max}}}\,,
\end{align}
where $L_J = \sqrt{(\pi c_s^2) / (G\rho)} {\Delta s}^{-1}$ is the number of cells per Jeans length, and $\Delta s$ is the cell size in the patch. 
These criteria prevent numerical fragmentation due to under-resolution of the Jeans length \citep{truelove} and promote refinement in shear flows.
The tunable parameters in our models are set to $L_{\rm J,min}=8$ and $\mathrm{\omega_{max}}=5$. 
To ensure some hysteresis between refinement and de-refinement, decisions are made based on the maximum value in the patch of the combined criteria as follows
\begin{equation}
1.4\, \left[R_J + R_\omega \right]_{\max} \left\{
\begin{array}{ll}
      < 0.4\,&:  \textrm{derefine; destroy patch} \\
      > 1\,&:  \textrm{refine; create child patches} \\
      \textrm{else}&:  \textrm{maintain patch}\\
\end{array} 
\right.
\end{equation}
The criterion is only evaluated every five timesteps to avoid situations where child patches could be repeatedly created and destroyed, allowing the dynamics to develop at the higher resolution, before derefinement is potentially triggered due to a smoother flow. Using combined criteria results in a smoother refinement ladder compared to using each criterion individually as a sufficient condition. Furthermore, the chosen parameters lead to generous refinement inside the geometric distance of 144 cells. We find that typically 43\,\% of the $7.6\,\times10^7$\,total cells are refined above level 19 (1.6\,au) in the zoom-in models.

\subsection{Parental run}
\label{sec:methods_parental_run}
The parental simulation of a molecular cloud environment has a (4\,pc)$^3$ cubic simulation domain (Fig.\,\ref{fig:panel}). It contains a total mass of $3000$\,M$_\odot$ and is initiated with a mean magnetic field of \SI{7.2}{\micro G}.
This simulation has also been used in prior works to investigate late-stage infall \citep{kuffmeier_2023}, chemistry within the core \citep{jensen_2021_corechemistry, jensen_2023_corechemistry}, 
accretion of mass and angular momentum through Bondi-Hoyle accretion in Class II systems  \citep{padoan_2025_BHaccretion,pelkonen_2025_angularmomentuminclassII}, and \textsc{ramses}-based zoom-in studies within molecular clouds \citep{jorgensen_2022_boxused, tuhtan_2023, Al_Belmpeisi_2024_simulatedanalogousII}.
Here, we only briefly describe the setup, which has been presented in detail for similar models in \citet{haugbølle_2018}.
The ideal MHD equations are solved using a $512^3$ root grid and 6 refinement levels, achieving a minimum cell size of 25\,au.
Starting with a uniform density and magnetic field, we apply a solenoidal random forcing on the largest scales for 20 dynamical timescales, defined as half the box size divided by the root mean square velocity.
This is done to erase any memory of the initial conditions and fully develop a turbulent cascade throughout the box with a sonic Mach number of 10.
Subsequently, self-gravity is turned on while maintaining the turbulent driving, leading to the formation of sink particles (Fig.\,\ref{fig:panel}).
During the full duration of the parental run simulation of $2.1$\,Myr, a total of 321 parent sinks are formed, with a sink formation efficiency of 8.5\,\%.  
When unclear, we denote the sink particles from the \textsc{ramses} parental run as ``parent sinks'' 
to differentiate them from the sinks in the \textsc{dispatch} re-simulations.

\subsection{Criteria for the selection of parent sinks}
\label{sec:sink_selection}
In the parent run, from the several hundred 
sinks, nine sinks were selected for re-simulation 
(parent sink: \texttt{6}, \texttt{13}, \texttt{14}, \texttt{25}, \texttt{82}, \texttt{122}, \texttt{162}, \texttt{180}, and \texttt{225}). We chose the nine parent sinks according to two overarching criteria: no other 
sinks form or come closer than $10^4$\,au, during the first 100 kyr and that the stellar mass is in the interval 0.5\,M$_\odot$\,to 1\,M$_\odot$\,at 100 kyr. These criteria were fulfilled by 30 candidates, from which we then selected 9 representative parent sinks. 
In general, the selected sinks represent some of the most isolated sinks with the least amount of interaction and dynamical variation in the inflow properties.
Properly modelling clustered star formation, with higher multiplicity and stellar interactions, would require a different zoom-in strategy \citep{jorgensen_2022_boxused,tuhtan_2023}.

\begin{table*}
    \centering
    \caption{\label{tab:core_overview} 
    Core properties measured within a $R=10^4$\,au radius from the central star. 
    Row 1 gives the global time for star creation.
    Rows 2 and 3 are the total mass and the average gas number density. 
    Row 4 gives the velocity dispersion.
    Rows 5 and 6 are the volume-averaged mean magnetic field and variability of the B-field. 
    Row 7 is the mass-to-flux ratio along the volume-average mean magnetic field direction, $\mu(\Vec{B}_\perp)$. 
    Row 8 is the angle between volume-average mean magnetic field direction and the total angular momentum vector for the core, $\theta(\Vec{B},\Vec{L})$. 
    Rows 9 to 13 are the kinetic, magnetic, thermal, rotational and gravitational energy. The final row is the ratio of rotational energy to gravitational energy, $\beta_{\rm rot}$.}
    \setlength{\tabcolsep}{6pt}
    \begin{tabular}{clc|ccccccccc|c}
    \hline\hline
     & \multirow{2}{*}{Quantity}    & \multirow{2}{*}{Unit}                 & \multirow{2}{*}{\texttt{6}}  & \multirow{2}{*}{\texttt{13}}    & \multirow{2}{*}{\texttt{14}}    & \multirow{2}{*}{\texttt{25}}    & \multirow{2}{*}{\texttt{82}}    & \multirow{2}{*}{\texttt{122}}   & \multirow{2}{*}{\texttt{162}}   & \multirow{2}{*}{\texttt{180}}    & \multirow{2}{*}{\texttt{225}}   & \multirow{2}{*}{Avg.}             \\[10pt]
    \hline
    1  & $T_0$                     & [Myr]                 & 0.15 & 0.24 & 0.24 & 0.32 & 0.56 & 1.0  & 1.2  & 1.3   & 1.6  & --              \\
    2  & $M$                       & [M$_\odot$]           & 1.82 & 1.87 & 1.94 & 2.09 & 1.78 & 1.83 & 1.68 & 1.47  & 1.66 & $1.8\pm0.2  $   \\ 
    3  & $\langle n\rangle$        & [10$^{3}$ cm$^{-3}$]  & 65   & 68   & 70   & 76   & 64   & 66   & 61   & 53    & 60   & $65\pm6     $   \\
    4  & $\sigma_v$                & [km s$^{-1}$]         & 0.32 & 0.36 & 0.32 & 0.41 & 0.34 & 0.36 & 0.27 & 0.32  & 0.36 & $0.34\pm0.04$   \\
    5  & $\langle B\rangle$        & [\SI{}{\micro G}]     & 41   & 51   & 33   & 94   & 68   & 50   & 51   & 31    & 62   & $53\pm20    $   \\
    6  & $\sigma_B$                & [\SI{}{\micro G}]     & 81   & 85   & 57   & 81   & 56   & 83   & 50   & 38    & 57   & $65\pm19    $   \\
    7  & $\mu(\Vec{B}_\perp)$      &  --                   & 2.57 & 2.19 & 3.12 & 1.28 & 1.51 & 2.01 & 1.88 & 2.67  & 1.49 & $2.1\pm0.6  $   \\
    8  & $\theta(\Vec{B},\Vec{L})$ & [deg]                 & 162  & 86.2 & 76.3 & 124  & 15.0 & 70.5 & 77.3 & 65.5  & 147  & $92\pm46    $   \\
    9  & $E_\text{kin}$            & [10$^{42}$ erg]       & 1.9  & 2.4  & 2.0  & 3.5  & 2.1  & 2.4  & 1.2  & 1.5   & 2.2  & $2.1\pm0.7  $   \\
    10 & $E_\text{mag}$            & [10$^{42}$ erg]       & 0.36 & 0.43 & 0.19 & 0.68 & 0.35 & 0.42 & 0.23 & 0.11  & 0.32 & $0.34\pm0.17$   \\
    11 & $E_\text{therm}$          & [10$^{42}$ erg]       & 1.2  & 1.4  & 1.4  & 1.4  & 1.2  & 1.2  & 1.1  & 0.97  & 1.4  & $1.25\pm0.16$   \\
    12 & $E_\text{rot}$            & [10$^{42}$ erg]       & 0.12 & 0.11 & 0.23 & 0.27 & 0.16 & 0.19 & 0.35 & 0.081 & 0.64 & $0.24\pm0.18$   \\
    13 & $|E_\text{grav}|$ & [10$^{42}$ erg] & 6.0 & 4.4 & 3.8 & 2.5 & 4.4 & 2.6 & 3.2 & 2.3 & 6.2 & $3.9\pm1.4$ \\
    14 & $\beta_{\rm rot}$ & -- & 0.019 & 0.025 & 0.061 & 0.11 & 0.036 & 0.072 & 0.111 & 0.035 & 0.10 & $0.063\pm0.035$ \\
    \hline\hline
  \end{tabular}
\end{table*}

\subsection{Re-simulation procedure in DISPATCH }

Following the selection of the  parent sinks, we took nine snapshots, prior to sink formation, from the parental run and converted them from the \textsc{ramses} format to the \textsc{dispatch} framework \citep{nordlund_2018, Ramsey_2018, popovas_2025}.
In this way, each {\sc dispatch} simulation inherits the full (4\,pc)$^3$ domain and retains a representative cloud environment, as opposed to artificially excising a reduced region. 
Using our zoom-in methodology, the resolution was then gradually increased from approximately $25$\,au to $0.8$\,au with a relaxation time corresponding to $l_\mathrm{p}/c_\mathrm{s}$, where $l_\mathrm{p}$ represents the characteristic length of each patch, and $c_\mathrm{s}$ the sound speed. 
The root grid level for the zoom-ins, level 7, has a resolution of $\approx$$6\times10^3$\,au, covering the largest structures of the environment. 
A representation of the applied zoom-in can be seen in Figure\,\ref{fig:zoomin}, showing the largest scales at 4 pc (panel a) and zooming in one order of magnitude per panel 
resulting in a box size of approximately 800 au in width in panel d of Figure\,\ref{fig:zoomin}.


\section{Natal environment}
\label{sec:resIC}
Below, we refer to sinks as “stars” and use “systems” to denote either the cores or the combined star–disc environments.
\subsection{Formation of stars in a molecular cloud environment}
\label{sec:resutls_initial_condition}

Stars form in dense filaments created by turbulent fragmentation. After formation, they move through the cloud in an ever-changing and dynamic environment.  
Figure\,\ref{fig:panel} displays three snapshots in time from the 
previously-described parental run (Sec.\,\ref{sec:methods_parental_run}).
The first generation of stars includes 7 stars seen in the top left part of 
panel a of Figure\,\ref{fig:panel}.
Another 116 stars emerge over the next $0.86$\,Myr, concentrated in the filamentary regions, seen in panel b of Figure\,\ref{fig:panel}. We remark that, due to the projection and the precision of the stellar positions, some stars appear to be overlapping. 
In the lower right corner of panels b and c (Fig.\,\ref{fig:panel}), some stars are less embedded, as they reside on the edge of low-density voids, transitioning into a Class II stage \citep{pelkonen_2025_angularmomentuminclassII}. 
The final stellar number density within the box is 5.0 pc$^{-3}$, 
corresponding to a mean distance between stars of $0.58$\,pc.

Zooming in on an individual system (\texttt{{13}}) reveals a star-forming filament, an infalling envelope, and an embedded disc (Fig.\,\ref{fig:zoomin}). 
A filament, approximately $0.05$\,pc in width, is shown in panel b of Figure\,\ref{fig:zoomin}. This scale corresponds to the Jeans length $\lambda_{\rm L}=0.12$\,pc in the higher density environments with $\rho=10^{-19}$\,g\,cm$^{-3}$ in our simulation domain \citep{jeans_1902}. 
This is broadly consistent with observed length scales of star-forming filaments \citep{pineda_2023,zhang_2024_obs_filaments}.
Panel c of Figure\,\ref{fig:zoomin} 
is a zoom-in on the $10^4$\,au-core scale and reveals the embedded star with an infalling envelope, measuring approximately $0.02$\,pc, or $4000$\,au, in size. 
A long vertical structure connecting the cloud environment to the disc can be seen in panel c.
The final panel d shows a $100$\,au-sized disc deeply embedded in a turbulent environment.

\subsection{Properties of selected cores}\label{sec:resutls_the_cores}

The initial prestellar core properties of the nine selected systems are listed in Table\,\ref{tab:core_overview}. The values are calculated at the star creation time (row 2 of Table\,\ref{tab:core_overview}). The right-outermost column displays the average value across all the cores, along with the corresponding variability ($\sigma_{\rm core}$), i.e. not the uncertainty on the mean. 
The core region is here defined ad hoc as a uniform sphere with a $10^4$\,au radius around the central star, providing a uniform scale to allow comparison of the environment around the selected newly-formed stars.

\begin{figure}
    \centering
    \resizebox{0.9\hsize}{!}{\includegraphics{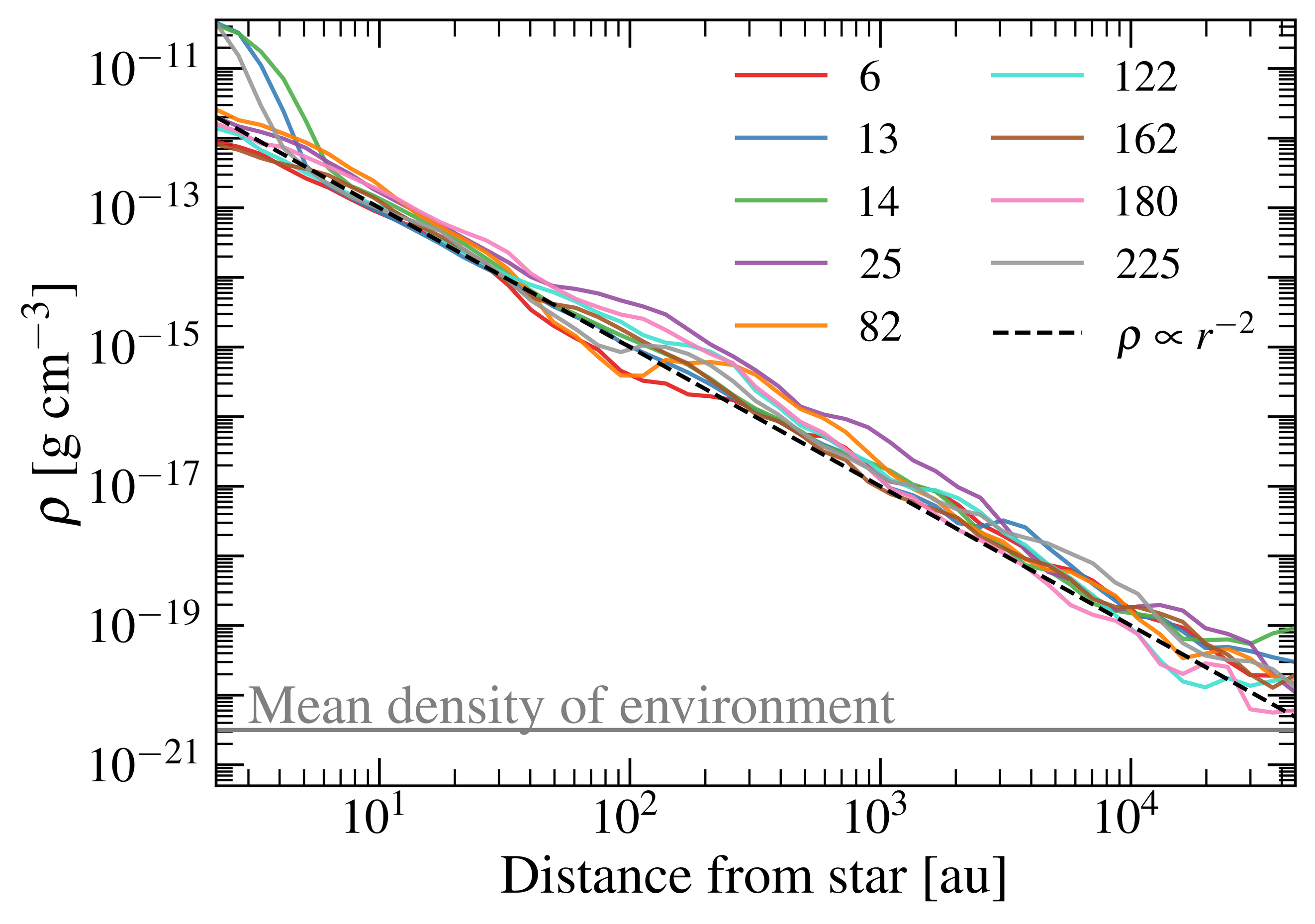}}
    \caption{\label{fig:core_densityprofile}
    Density profiles of the nine selected systems at the star creation time $T_{0}$ (see row 2 in Table\,\ref{tab:core_overview}).
    The density profile of a theoretical singular isothermal sphere is shown with the black dashed line. 
    The mean density of the full environment is shown with the horizontal grey line.}
\end{figure}

\subsubsection{Mass budget}
The selected cores have an average density of $\langle\rho\rangle = 2.55\times10^{-19}$\,g\,cm$^{-3}$
and an average mass of $\langle M \rangle=1.8\pm0.2$\,M$_\odot$ within $10^4$\,au.
If we instead consider a core radius where the gravitational energy dominates over the bulk kinetic and thermal energy (virial number $\alpha_{\rm vir}<1$, Fig.\,\ref{fig:non-alpha_vir}), we find core masses in the range $0.1$ -- $1.1$\,M$_\odot$.
However, given the turbulent and dynamic environment, there is no strong correlation between initially bound mass and final accreted mass for a single system \citep{pelkonen_2021_unboundaccreting_material, kuffmeier_2023, kaalva_2026_boundcore}.

The radial density profiles of the cores are shown in Figure\,\ref{fig:core_densityprofile}. Despite the spread in the core properties listed in Table\,\ref{tab:core_overview}, their density profiles are well approximated by a singular isothermal sphere with density scaling as $\rho\propto r^{-2}$ \citep{ebert_sphere1955, bonnert_sphere1956}. 
Similar core density profile was reported for star \texttt{13} in \citet[][their Fig.\,7]{tuhtan_2023}. 
Notably, in \citet[][their Fig.\,2]{mayer_2025_largescale_niMHD_structures} the core density profiles show stronger deviations from the spherical core profile compared to our results (Fig.\,\ref{fig:core_densityprofile}).
Our measured deviations from the power-law profile are only noticeable in the innermost part of the core, indicating that the collapse has already begun. At distances exceeding $10^4$\,au, core densities approach the mean background density of the simulation domain, $\rho=3.2\times 10^{-21}$\,g\,cm$^{-3}$. 
This background density can be compared to the higher background densities explored in previous studies of up to $2\times 10^{-18}$\,g\,cm$^{-3}$ \citep{seifried_2013_turbulence_and_Bfield}, $1.2\times10^{-19}$\,g\,cm$^{-3}$ \citep{Bate_2018}, and $1.5-3.0\times10^{-19}$\,g\,cm$^{-3}$ \citep{lebreuilly_2024a}. Such values are more in line with the densest regions of molecular clouds, whereas our background density corresponds to the average densities of molecular clouds with column densities of $N_{H_2}\approx 10^{22}$\,cm$^{-2}$ \citep{andre_2014_nonspherical_cores}.

\subsubsection{Kinetic properties of the cores}

The mass-weighted velocity dispersion, $\sigma_v$, for the nine cores ranges from 0.27 km s$^{-1}$ to 0.41 km s$^{-1}$, with an average of $\langle \sigma_v\rangle =0.34\pm0.04$  km s$^{-1}$. 
This is consistent with observed mean velocity dispersion, $\sigma_v=0.29$\,km\,s$^{-1}$, across 288 low-mass cores ($<10$ M$_\odot$) in the ASHES survey \citep{Li_2023_velocity_dispersion}.

The cores are dominated by kinetic energy, pointing to the inherited turbulence from the parent molecular cloud. 
Table\,\ref{tab:core_overview} lists the total kinetic energy, as well as the thermal energy.
The final rows give the total rotational, gravitational energy, and the ratio of these, $\beta_{\rm rot}$. The rotational energy is calculated following \citet{chen_2018_rotational_energy}, showing a rotational to kinetic energy ratio of $E_{\rm rot} / E_{\rm kin}\sim0.1$, in line with the ideal MHD core simulations of 
\citet{chen_2018_rotational_energy}. The final row 14 contains $\beta_{\rm rot}$, defined as $\beta_{\rm rot}=E_{\rm rot}/E_{\rm grav}$ and are comparable to the ones presented in \citep{wurster_bate_2019} and the canonical value of $\sim0.02$ \citep{pineda_2023}.

\subsubsection{Magnetic properties of the cores}
\label{sec:Bcore}

The volume-averaged mean magnetic field strength and magnetic field dispersion of the nine cores are $\langle \langle B \rangle \rangle=$ $53\,\pm$\,\SI{20}{\micro G}
and $\langle \sigma_B \rangle=65\,\pm$\,\SI{19}{\micro G}. These values fall within the range inferred from Zeeman observations of dense molecular cores, which typically show field strengths of 10-\SI{100}{\micro G} 
with only a weak dependence on density
\citep{crutcher_2010_meanfieldincores, Crutcher_2012_first_Bfield_in_core}.
Row 10 in Table\,\ref{tab:core_overview} shows the total magnetic energy $E_\text{mag} = B^2/8\pi$ for each core.
We note, that the cores reported in \citet{mayer_2025_largescale_niMHD_structures} are more strongly
magnetised, having average magnetic strengths and dispersion one order of magnitude higher than ours.

The mass-to-flux ratio measures the relative importance of gravity to magnetic support, and is defined as
\begin{equation}
    \mu = \frac{M_{\text{core}}}{\Phi_{\text{core}}} \Bigg/ \left( \frac{M}{\Phi} \right)_{\text{crit}}\,.
    \label{eq:mass2flux}
\end{equation}
Here, $M_\mathrm{core}$ is the mass of the core and $\Phi_\mathrm{core}$ the magnetic flux through a surface perpendicular to the volume-averaged magnetic field vector.
We use $(M/\Phi)_\mathrm{crit} = 1/\sqrt{63G}  \approx 0.13/\sqrt{G}$ as the critical value for the collapse of a magnetised core \citep{mouschovias_1976_mass2flux}, which is close to $(M/\Phi)_\mathrm{crit} =1/(2\pi\sqrt{G})\approx 0.16/\sqrt{G}$ obtained from stability analysis of a magnetized sheet \citep{nakano_1978_sheet, kunz_2009_masstoflux_core}. Here, $G$ is the gravitational constant.
The average mass-to-flux ratio across the cores is $\langle \mu \rangle=2.1\pm0.6$. 
This is significantly lower than $\mu=23\pm16$ reported in the work by \cite{pelkonen_2021_unboundaccreting_material}, which used a model with the same parameters but lower resolution than our parental run, because we compute the magnetic flux along the volume-averaged magnetic field direction, yielding a higher flux, as opposed to planes defined by the global simulation axis (see Appendix\,\ref{sec:appendix_masstoflux} for a further description). 

In our sample, the inclination between spin axis, $\Vec{L}$, and the volume-average B-field direction shows large misalignments on scales of $10^4$\,au (row 8, Table\,\ref{tab:core_overview}).  
This also holds true on scales of $10^3$\,au (Table\,\ref{tab:mass_to_flux_overview}), much closer to the central star, with only the \texttt{180} core showing near-alignment $\theta(\Vec{B},\Vec{L})\approx 20^\circ$. Table\,\ref{tab:mass_to_flux_overview} in Appendix\,\ref{sec:appendix_masstoflux} gives a full overview of 
the angle between the mean magnetic field direction and total angular momentum vector, the mass-to-flux measured for three planes associated with the magnetic field direction and the total angular momentum vector. This is presented for spheres with radii $R=10^3$\,au and $R=10^4$\,au.


\section{Disc formation and evolution}\label{sec:disc_formation_evolution}
\label{sec:discformevol}
\begin{figure*}[t!]
    \centering
    \resizebox{\hsize}{!}{\includegraphics{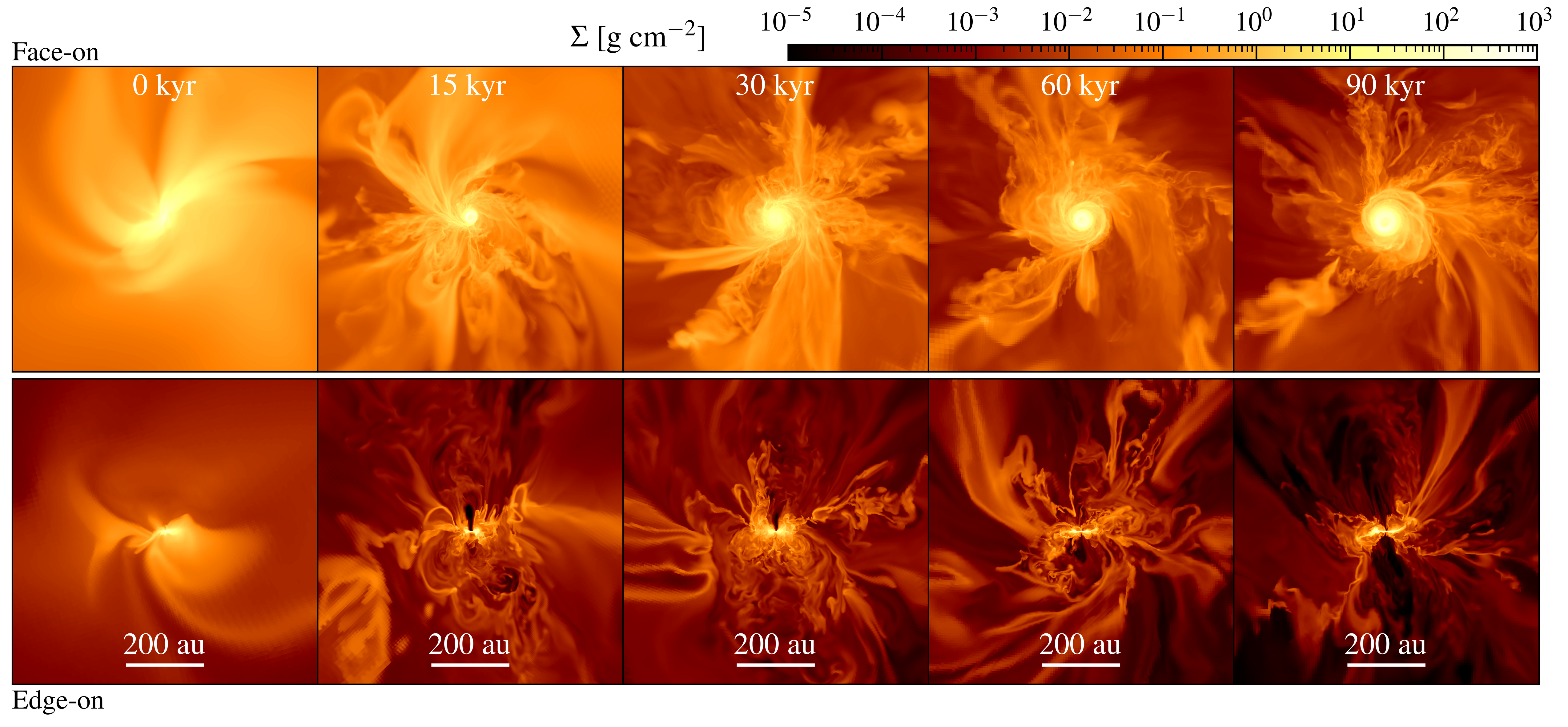}}
    \caption{
    Time evolution of face-on (top panels) and edge-on (bottom panels) column densities for system \texttt{180}.
    The integration depths are $80$\,au for the face-on 
    panels
    and $5$\,au for the edge-on panels, in order to best visualise the empty polar cavities made by outflows. 
    The disc grows with time through filamentary infall, depleting the surroundings of gas.
    }
    \label{fig:osyris_rho_7panel_s180cm5}
\end{figure*}

In this section, we 
characterise the formation and evolution of 
the star-disc environment of system \texttt{180}.
This system was selected because it did not experience a major disruptive streamer event.
Moreover, the inner part of the core ($<10^3$ au) shows near alignment 
between spin-axis and volume-average mean magnetic field 
($\theta( \Vec{B}, \Vec{L}) \approx 20^\circ$, Appendix\,\ref{sec:appendix_masstoflux}), which facilitates the analysis. 

\subsection{Overview of the formation and evolution of system 180}

Discs emerge and grow in mass and size over time through filamentary transport of material to the system, while shaping the environment by creating low-density polar cavities through outflows.
This process is illustrated in Figure\,\ref{fig:osyris_rho_7panel_s180cm5}, which shows the integrated density, along edge- and face-on orientations with respect to the forming disc.
Around a newly-formed star ($t=0$\,kyr) no disc has emerged yet.
However, within $15$\,kyr, a small rotationally-supported disc has developed (top $15$\,kyr-panel), with an approximate radius of $10$\,au. 
At $t=30$\,kyr, three prominent streams are supplying material to the disc (top  $30$\,kyr-panel). Similar filamentary structures can be seen at later times.
For example, two noticeable streams are present in the upper-right and lower-left quadrants of the final $90$\,kyr edge-on panel.
In this way, the disc around star \texttt{180} grows in size, developing an inner high-density region within approximately $50$\,au radius and a less-dense peripheral area extending out to 100 au radius (top  $90$\,kyr-panel). 
We analyse in detail the mass transport towards the disc (Sec.\,\ref{sec:results_flow_and_transport}), the corresponding magnetic field evolution (Sec.\,\ref{sec:magnetic_fields}), and the structure and evolution of the disc itself (Sec.\,\ref{sec:results_surfac_density} -- \ref{sec:results_mass_transport}). 

The density in the region surrounding the disc gradually decreases over time, a trend particularly evident in the edge-on 
panels of Figure\,\ref{fig:osyris_rho_7panel_s180cm5}. A low-density cavity is already visible in the bottom panels after $t\gtrsim 15\,$kyr.
This cavity is created by an outflowing stream of material.
These outflows are generally intermittent, asymmetric and misaligned with the polar axis due to the highly turbulent surroundings \citep{gerrard_2019_misalignedjets}.
The cavities above and below the midplane do not appear in conjunction, signifying asymmetric outflows seen both in observations \citep{Wassell_2006_asymmetric_outflows_obs} and models \citep{Dyda_2015_asymmetric_outflows_model}. 


\subsection{Characterising the mass flux onto a newly-formed disc}\label{sec:results_flow_and_transport}

\begin{figure*}
    \centering
    \resizebox{\hsize}{!}{\includegraphics{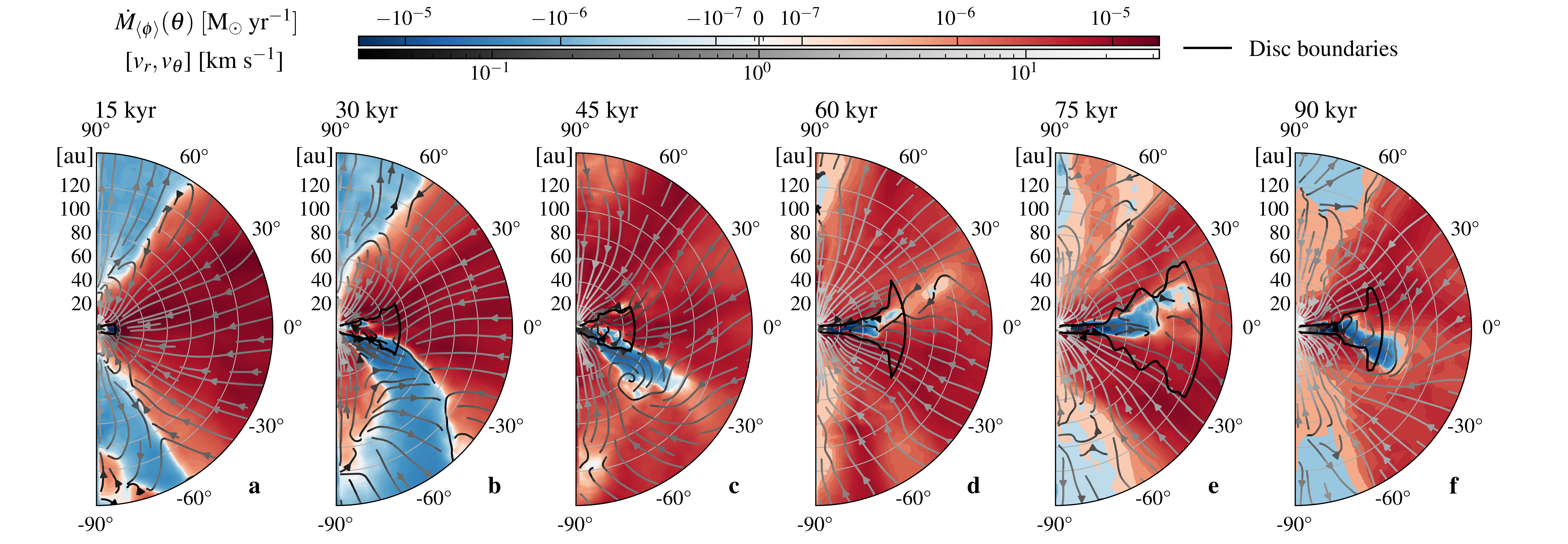}}
    \caption{Time and azimuthally averaged mass flux $\dot{M}_{\langle\phi\rangle}$, with the poloidal component of the flow. The spin axis is defined as the total angular momentum vector within 150 au. All values are averages over 1.1 kyr, corresponding to the dynamical timescale of the disc. The streamlines represent the poloidal component of the velocity, i.e. $v_r$ and $v_\theta$. The black line shows the fitted scale height and the disc edge, together making up the disc boundaries. The definition of the disc boundaries will be presented in Section \ref{sec:results_surfac_density}. The disc undergoes a formation epoch ($t<50$\,kyr) dominated by polar outflows and strong accretion, followed by an evolution phase ($t>50$\,kyr) characterised by mass-flux along the high-density surface-layer of the disc.}
    \label{fig:azimuthal_6avg_dmdtheta_150au_s180cm5}
\end{figure*}

\begin{figure}
    \centering
    \resizebox{\hsize}{!}{\includegraphics{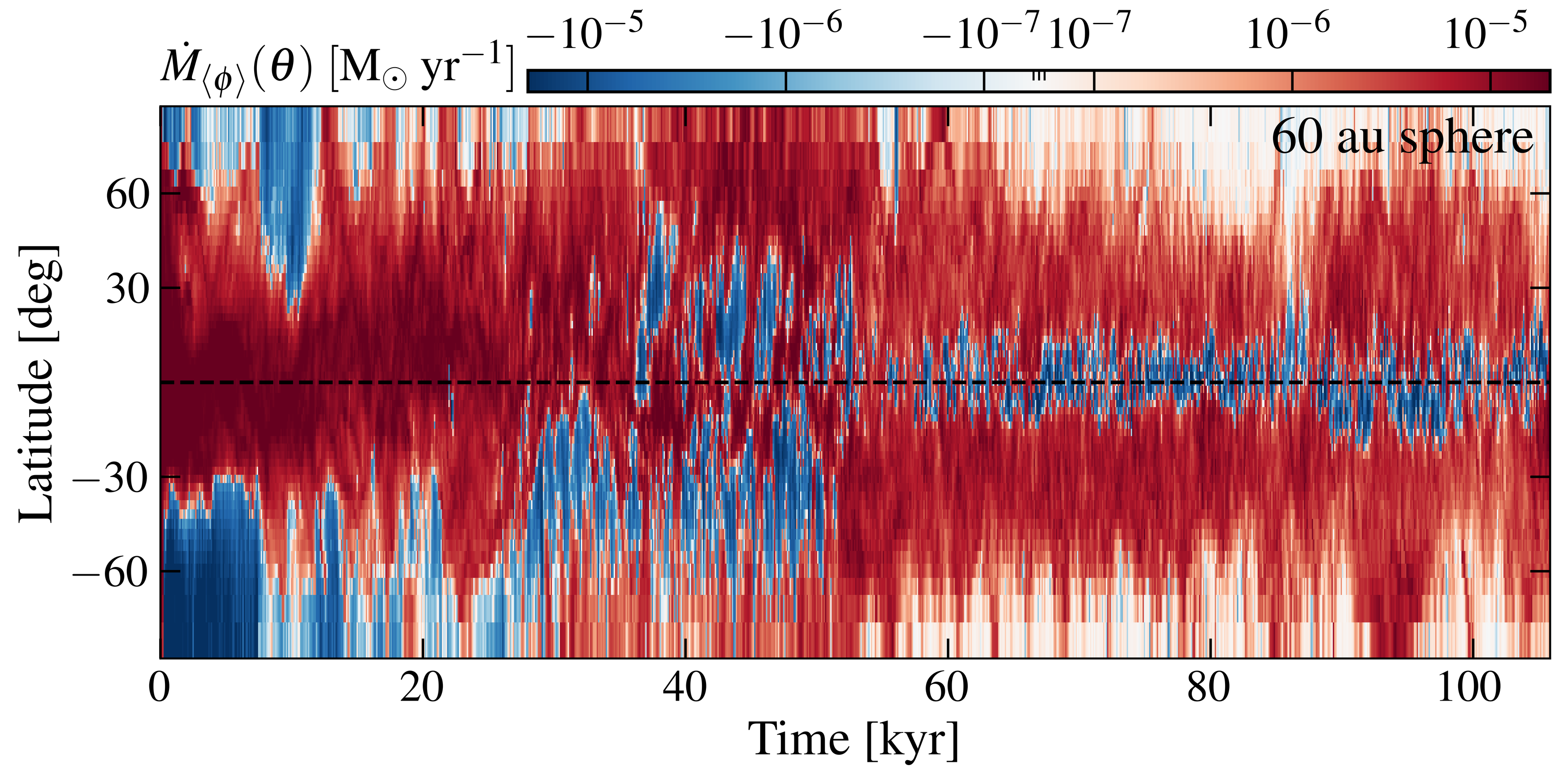}}
    \caption{Full time evolution of the mass flux through a 60 au shell for system \texttt{180}, showing ${\dot{M}_{\langle\phi\rangle}}$ as a function of latitude with respect to the midplane and time. Blue signifies outflow and red inwards accretion.}
    \label{fig:butterfly_60au_s180cm5}
\end{figure}

The nature of the mass flux of gas, inwards from the surrounding environment down to both the disc and the central star, evolves with time.
We define the radial mass flux as a function of polar angle as the azimuthal average of the mass flux, normalised to the full sphere,
\begin{equation}
    \dot{M}_{\langle\phi\rangle}(\theta) = \mathbf{-}4\pi r^2 \langle\rho v_\mathrm{r}\rangle_{\phi}.
    \label{eq:massflux_per_unit_polarangle}
\end{equation}
Here, $\rho$ is the gas density, $v_\mathrm{r}$ is the net radial component of the velocity, and $\theta$ is the polar angle measured from the midplane.
Figure\,\ref{fig:azimuthal_6avg_dmdtheta_150au_s180cm5} shows $\dot{M}_{\langle\phi\rangle}(\theta)$ at different times for system \texttt{180}. Time averaging is applied to smooth-over turbulent fluctuations on a timescale of $10^3$\,yr. This makes it possible to discern trends in the accretion pattern, but we stress that the non-smoothed dynamics is highly variable and turbulent. 

%
Substantial bipolar outflows are launched almost immediately after the formation of the host star (Fig.\,\ref{fig:azimuthal_6avg_dmdtheta_150au_s180cm5}, panel a). These outflows, launched close to the host star with initial outflow velocities reaching up to $10$\,km\,s$^{-1}$, carry an outward mass flux on the order of $\sim$10$^ {-6}$\,M$_\odot$\,yr$^{-1}$.
This bipolar outflow limits the inward flow of gas to a polar opening angle of $-60^\circ\lesssim\theta_{\rm acc}\lesssim60^\circ$ with respect to the disc midplane (Fig.\,\ref{fig:azimuthal_6avg_dmdtheta_150au_s180cm5}, panel a). 
%
Within this angle of accretion, $\theta_{\rm acc}$, most of the material flows directly to the star-disc system at an accretion rate of $\dot M$\,$\sim$ 10$^{-5}$ M$_\odot$\,yr$^{-1}$. 
An important caveat is that our 0.8\,au resolution is insufficient to resolve high-velocity jets 
originating from close ($\lesssim 0.1$\,au) to the host star \citep{tu_2026_highvelocityjets}.
We also do not employ a jet prescription \citep{federrath_2014_subgrid_jets} in our sink model (Appendix\,\ref{appendix:sink_cell_prescription}).

Later, at $t=30$\,kyr, the cavities created by the outflows are still pronounced in both polar regions (Fig.\,\ref{fig:azimuthal_6avg_dmdtheta_150au_s180cm5}, panel b).
However, in the southern hemisphere, intermittent and off-axis outflows allow for a small region of infall 
along the polar axis, limiting inwards accretion of gas to an opening angle of approximately $-30 ^\circ\lesssim\theta_{\rm acc}\lesssim60 ^\circ$.

The emerging rotationally-supported disc is shown with a black contour in Figure\,\ref{fig:azimuthal_6avg_dmdtheta_150au_s180cm5} marking the vertical gas scale height and the outer disc edge (defined in Section\,\ref{sec:results_surfac_density}). 
Whereas midplane accretion is at first dominantly inwards-directed (Fig.\,\ref{fig:azimuthal_6avg_dmdtheta_150au_s180cm5}, panels a-c), at later times ($t \gtrsim 50$\,kyr) the midplane shows also outwards mass flows
(Fig.\,\ref{fig:azimuthal_6avg_dmdtheta_150au_s180cm5}, panels d-f). 

The last three panels d, e, and f of Figure\,\ref{fig:azimuthal_6avg_dmdtheta_150au_s180cm5} show the establishment of a near steady-state layered accretion structure after 
$t \sim 50$\,kyr. While the disc midplane ($|\theta | \lesssim 30^\circ$) exhibits a highly turbulent flow, the disc surface ($30^\circ\lesssim | \theta_{\rm acc} | \lesssim 50^\circ$) is dominated by inwards accretion flows. This mode of layered surface accretion is consistent with previous ideal MHD global disc simulations \citep[][]{suzuki_2014_surface_accretion,zhu_stone_2018} and the resistive MHD 2D simulation by \citet{suriano_2017_Bfield_transport_layer}, further described in Section\,\ref{sec:magnetic_fields}. 
The polar regions ($ |\theta_{\rm} | \gtrsim 60^\circ$) have a low outwards mass flux of $\sim$10$^{-6}$--10$^{-7}$ M$_\odot$ yr$^{-1}$.
Note that due to the azimuthal averaging, low-density flows become more challenging to trace, causing narrow off-axis outflows to be smoothed out and potentially lost in projection. 

In summary, we identify a transition from a disc formation stage to a disc evolution stage, after approximately $t$$\approx$$50$\,kyr, where a rotationally supported accretion disc has formed, and accretion onto the host star predominantly occurs along the disc surface.
This transition is further illustrated in Figure\,\ref{fig:butterfly_60au_s180cm5}, showing the time evolution of the accretion rate through a shell with $r=60$\,au as a function of polar angle and time.\footnote{The shell values are projected from a volume of 60 $\pm$ 6 au, down to equal area cells in \textsc{Healpy}. From this, we calculate the average cell values binned to 51 equal-area latitude bands, accounting for the otherwise skewed distribution of latitudes.} 
During disc formation ($t \lesssim 50$\,kyr), accretion to the star happens within $\pm$\,60\,$^\circ$, at a rate on the order of $\dot M \sim 10^{-5}$\,M$_\odot$\,yr$^{-1}$.
The polar cavities can be characterised as low-density regions with predominantly outflowing material at $\sim$ 10$^{-6}$--10$^{-7}$\,M$_\odot$\, yr$^{-1}$, throughout most of the evolution.
After $t\approx 50$\,kyr, a surface layer ($30^\circ\lesssim | \theta_{\rm acc} | \lesssim 50^\circ$) develops, as a wedge between the low-density outflowing cavities in the polar regions and the midplane, that drives inwards mass transport (red coloured region in Fig.\,\ref{fig:butterfly_60au_s180cm5}).

\begin{figure*}
    \centering
    \resizebox{\hsize}{!}{\includegraphics{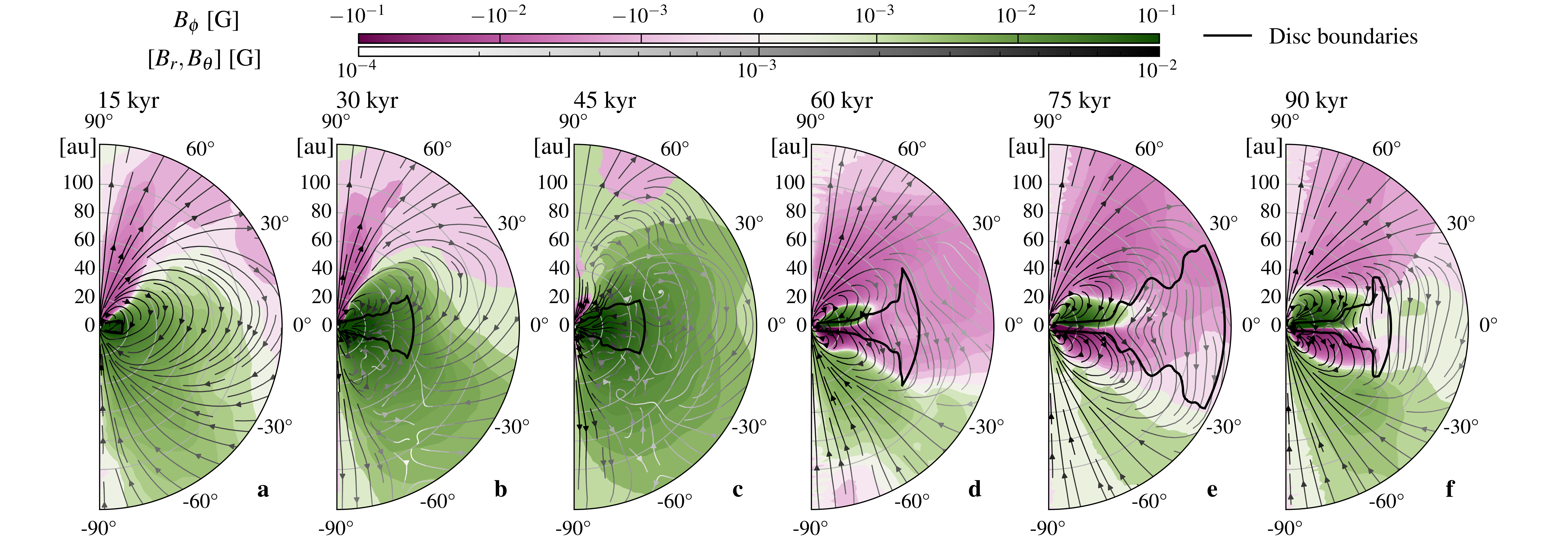}}
    \caption{The toroidal and poloidal field of the magnetic field is plotted here at 6 times, averaged over the dynamical scale of the disc ($1$\,kyr).
    The streamlines show the azimuthally-averaged poloidal field, 
    and the underlying colour map shows the toroidal component. 
    Note the difference in colour scale range between poloidal and toroidal fields.
    }
    \label{fig:azimuthal_6avg_120au_B_phi_Bpol_s180cm5}
\end{figure*}


\subsection{Evolution of the magnetic field structure}\label{sec:magnetic_fields}

The configuration of the magnetic field
evolves with time. During disc-formation ($t\lesssim 50$\,kyr) a near uniform positive toroidal magnetic field threads the disc.
Later evolution ($t\gtrsim 50$\,kyr) shows the winding up of an azimuthal magnetic field and a reversal across the disc midplane.
The full evolution of the magnetic field of system \texttt{180} is shown in Figure\,\ref{fig:azimuthal_6avg_120au_B_phi_Bpol_s180cm5}. 

The formation of system \texttt{180} occurs in an environment with relatively low magnetic field dispersion $\sigma_B$, compared to the other selected cores (Sec.\,\ref{sec:Bcore}).
Moreover, the small angle between the B-field and spin axis $\theta(\Vec{B},\Vec{L})\approx 20^\circ$ (Appendix\,\ref{sec:appendix_masstoflux}) gives rise to a relatively ordered and coherent initial poloidal field structure.
During the earliest disc growth stages ($t\lesssim 15$\,kyr), a magnetic dipole threads the disc midplane (Fig.\,\ref{fig:azimuthal_6avg_120au_B_phi_Bpol_s180cm5}, panel a).
We do not assume here a stellar dipole magnetic field model on the central star \citep{bouvier_2007_stellardipole}. 
Instead, we have verified that this dipole configuration already emerged in the contracting turbulent cloud material prior to the formation of the protostar ( $\sim$$1$\,kyr before $T_0$) with a mass-to-flux ratio of $\mu(\Vec{B}_\perp)=3.7$ within a volume of radius 80\,au.  The possibility of such dipole magnetic fields, as opposed to the classical hourglass-shaped split monopole field, due to turbulent reconnection, has been previously discussed as a pathway to prevent too efficient angular momentum loss during disc formation by partly decoupling the disc from the surrounding environment \citep{gonzalez_2016_progenic_dipole}.

As the disc grows in radius (Fig.\,\ref{fig:azimuthal_6avg_120au_B_phi_Bpol_s180cm5}, panels a-c), it starts to wrap up a toroidal field that dominates, with strength $|B_{\rm \phi}| \approx 0.1$\,G, over the poloidal component with strength $|B_{\rm pol}| \approx 0.01$\,G.
However, at later times ($t \gtrsim 50$\,kyr), we observe a transition to a significantly different magnetic field configuration,
corresponding to the transition from a disc formation stage to a more quasi-steady disc evolution stage (Sec.\,\ref{sec:results_flow_and_transport}).
In this latter stage, we observe that in the upper hemisphere the $B_\phi$ component is positive (along disc rotation) between $\theta\approx 0^\circ$ and $\theta\approx30^\circ$, while at higher latitudes ($\theta\gtrsim 30^\circ$) the $B_\phi$ component is negative (Fig.\,\ref{fig:azimuthal_6avg_120au_B_phi_Bpol_s180cm5}, panels d-f).
This change occurs across the high-mass-flux surface layer of the disc (see Fig.\,\ref{fig:azimuthal_6avg_dmdtheta_150au_s180cm5}, panels d-f). 
Also across the disc midplane ($\theta\approx 0^\circ$), the magnetic field flips sign from positive ($\theta\gtrsim 0^\circ$) to negative ($\theta\lesssim 0^\circ$), placing a current sheet directly in the midplane. The toroidal field alters back to positive across the lower disc surface ($\theta \lesssim -30^\circ$), producing a reverted azimuthal field symmetry across the midplane. 
The quasi-steady state of the gas flow field in the disc evolution stage after $t\approx 50$\,kyr, as discussed in Sec.\,\ref{sec:results_flow_and_transport}, is thus paired with a similarly stable magnetic field configuration. 

The magnetic field morphology seen here is consistent with the large-scale $B$-field structure seen in other ideal MHD simulations of isolated discs threaded by an imposed weak vertical magnetic field \citep{suzuki_2014_surface_accretion,zhu_stone_2018}.
Mass accretion occurs along the disc surfaces enclosing a turbulent midplane, pinching the poloidal component inwards, resulting in an azimuthally stretched field with sign reversal \citep{zhu_stone_2018}.
We recover this mode of surface accretion in our simulations of embedded discs, although in a lower plasma beta regime (plasma-$\beta$ shown in Figure\,\ref{fig:s180cm5_plasma_beta}). 
In contrast, we do not clearly see any intermittent disc winds \citep{suzuki_2014_surface_accretion,zhu_stone_2018}, which we suspect are suppressed due to the continuous infalling material on these young embedded discs.  
Recent work argues a similar mode of surface accretion is also prevalent on much smaller scales close to the star \citep[$\lesssim 0.1$\,au,][]{zhu_2023_global3dsimulationmagnetospheric}, but this is beyond the resolution limit of this study.


\subsection{{Disc analysis}}\label{sec:results_surfac_density}
  
\begin{figure}
    \centering
    \resizebox{0.9\hsize}{!}{\includegraphics{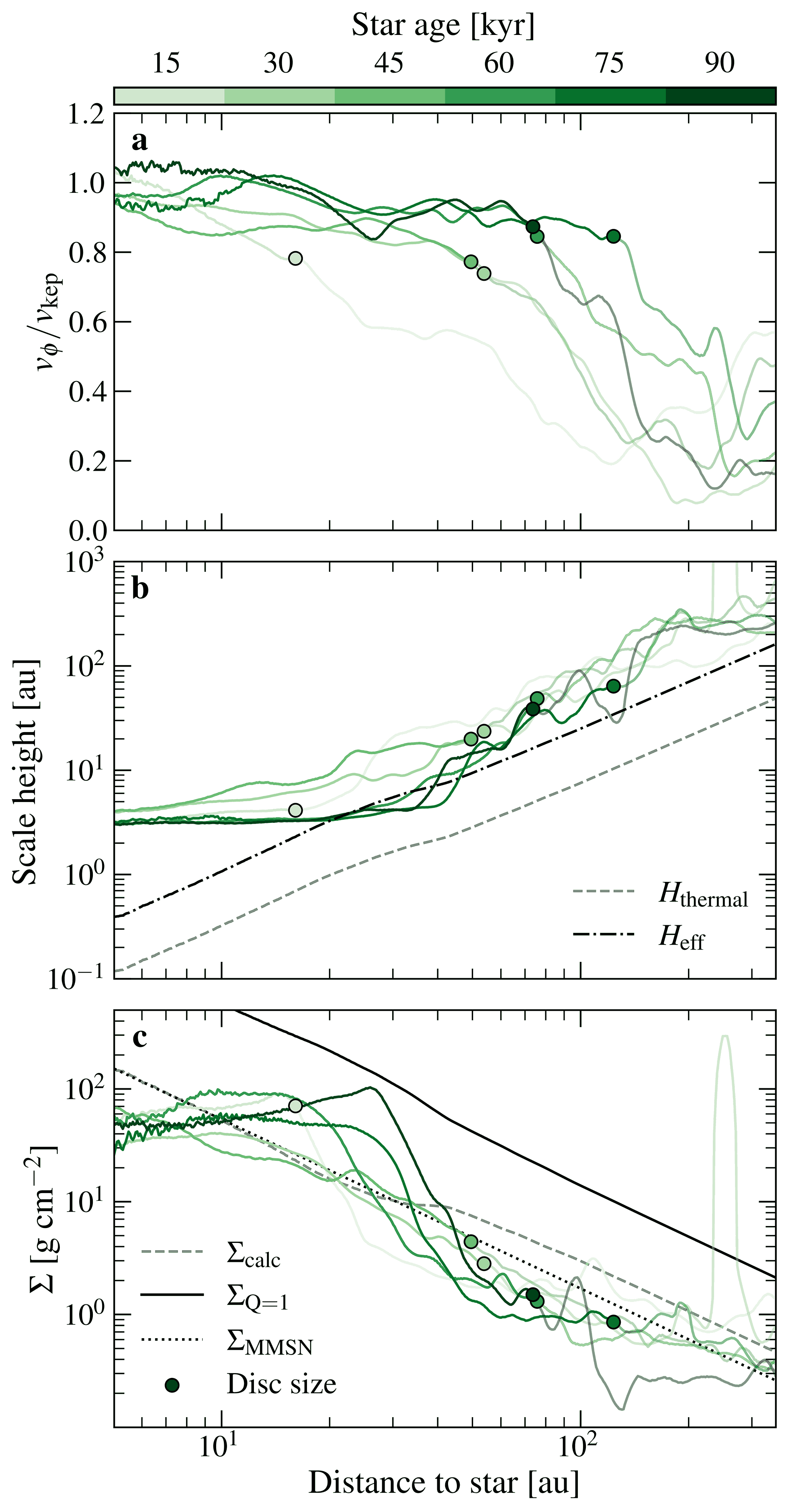}}
    \caption{
    Panel a: Time evolution of the toroidal velocity over true Keplerian velocity, which is one part of the definition of the disc size.
    Panel b: Time evolution of the gas scale height. 
    Panel c: Time evolution of surface density of system \texttt{180}. The dashed line shows the MMSN radius scaling as reference \citep{hayashi_1981_mmsn}. Here, $\varSigma_{\rm calc}$, is the calculated $\varSigma$ obtained from Eq.\,(\ref{eq:surface_density_approximation}) and plotted with the dashed dark green line. The surface density at 90 kyr if the disc is marginally gravitationally stable, denoted $\varSigma_{\rm Q=1}$, is plotted as solid dark line. 
    }
    \label{fig:s180cm5_gendisk}
\end{figure}

\begin{figure}
    \centering
    \resizebox{0.9\hsize}{!}{\includegraphics{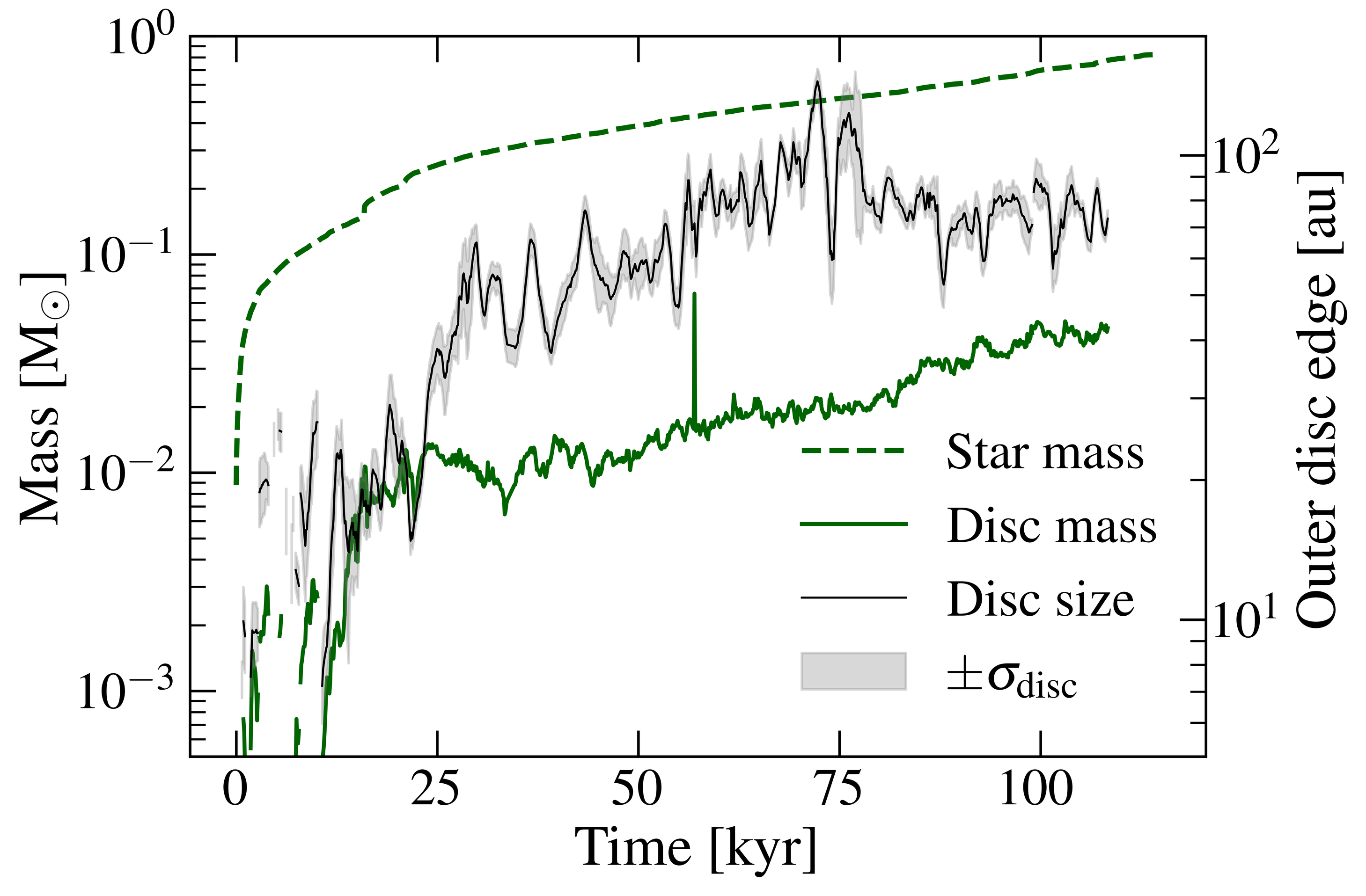}}
    \caption{High resolution time evolution of disc mass and size for system 180. The uncertainty derived from the disc size determination methods is plotted as the grey
    area. The disc sizes and masses are time averaged according to the dynamical timescale}
    \label{fig:disk_sizeandmass_s180cm5}
\end{figure}

\subsubsection{Evolution and determination of the disc radius}
\label{sec:results_disc_size}

We find that discs generally grow in radius with time. However, the outer edge of the disc does not represent a sharp density transition (see Sec.\,\ref{sec:results_surface_density}). Instead, discs remain connected to the surrounding material and disc edges are therefore hard to uniquely determine. 
Previous studies have therefore relied on arbitrary density thresholds \citep{joos_2012_magnetic_breakning_catastrophe, seifried_2013_turbulence_and_Bfield, lebreuilly_2024b, yang_2025} or various gas kinematics criteria \citep{machida_2014_discsizefromgaskinematics, kuffmeier_2017, tuhtan_2023}.
Instead, we combine two physical criteria to determine the disc radius. The first condition expresses that the disc is in near-Keplerian rotation,  
$v_\phi > 0.8\,v_{\rm kep}$,
where $v_{\rm kep}$ is the Keplerian velocity and $v_\phi$ is the mass-weighted azimuthal velocity (Fig.\,\ref{fig:s180cm5_gendisk}, panel a).
The second condition is the Rayleigh stability criterion for differentially rotating fluids. We give a full description of this new procedure in Appendix\,\ref{app:radfit}.
We show the ratio of orbital velocity and true Keplerian speed in panel a of Figure\,\ref{fig:s180cm5_gendisk} and mark the outer disc edge with a dot. 
Within this outer disc radius, gas generally rotates with a near-Keplerian velocity, while outside we see a steep decrease of the azimuthally-averaged velocity.

The full time evolution of the disc radius is shown in Figure\,\ref{fig:disk_sizeandmass_s180cm5} (black line).
After an initial period of roughly $10$\,kyr, a stable disc emerges and remains present thereafter. The initial disc size is approximately $15$\,au with high variability and fluctuations over time. 
Subsequently, the disc rapidly grows in radius to $r_{\rm disc} \approx 60$\,au at $40$\,kyr.
After this period of disc formation, the radius of the disc fluctuates between approximately $60$ and $100$\,au, stabilising to around $r_{\rm disc} \approx 80$\,au in the last $30$\,kyr of simulation time.

\subsubsection{Evolution of the gas scale height}\label{sec:results_scale_height}
Panel b of Figure\,\ref{fig:s180cm5_gendisk} illustrates the time evolution of the gas scale height, showing a gradual decrease with time as the host star grows in mass and vertical gravity increases. 
In the outer parts, the gas scale height has a near-constant aspect ratio $H/r \approx 0.2$.
Closer to the host star ($r\lesssim 20$\,au), scale heights are nearly constant with radius and range between $3$ and $10$\,au.
To obtain the gas scale height, we have used here a joint fitting procedure for both the gas scale height and surface density out to $r=500$\,au (Appendix\,\ref{sec:appendix_disc_fitting_procedure}), which is independent of the disc-radius criterion of Sec.\,\ref{sec:results_disc_size}.

Assuming vertical hydrostatic balance in the disc, the gas scale height scales as 
\begin{align}
    H_\mathrm{eff} 
    &= H_\mathrm{therm}\left( 1+ \frac{P_{\rm mag}}{P}\right)^{1/2} 
    = H_\mathrm{therm}\sqrt{1+\beta^{-1}}\,.
    \label{eq:Heff}
\end{align}    
Here, $P$ and $P_{\rm mag}$ are the thermal and magnetic pressure contributions. 
The thermal scale height is $H_\mathrm{therm}=c_{\rm s}/\varOmega_{\rm kep}$, where $c_{\rm s}$ is the local sound speed and $\varOmega_{\rm kep}$ the Keplerian frequency. 
The ratio of thermal to magnetic pressure is expressed by $\beta$. We find the thermal pressure to be subdominant in our barotropic ideal MHD simulations (Appendix\,\ref{appendix:Eos}), consistent with the finding of \cite{yang_2025}. 
The dashed grey line in panel b of Figure\,\ref{fig:s180cm5_gendisk} shows the thermal scale height 
for a nearly isothermal disc ($H_\mathrm{therm} \propto r^{3/2}$). The black dash-dotted line shows $H_\mathrm{eff}$ for a constant $\beta=0.1$, which is roughly consistent with the plasma-$\beta$ values in our simulations (Fig.\,\ref{fig:s180cm5_plasma_beta}) and approximately reproduces the measured scale height in the outer disc. However, in the inner disc ($r\lesssim 20$\,au) we develop a strongly magnetically-dominated
disc (Fig.\,\ref{fig:s180cm5_plasma_beta}). This is likely the result of efficient magnetic field transport in the ideal MHD limit towards the inner disc (Fig.\,\ref{fig:azimuthal_6avg_120au_B_phi_Bpol_s180cm5}) and, closer to the central star ($r<8$\,au), efficient removal of gas (Appendix\,\ref{appendix:sink_cell_prescription}).

\subsubsection{Evolution of the gas surface density and disc mass}\label{sec:results_surface_density}\label{sec:results_disc_mass}

\begin{figure*}
    \centering
    \resizebox{0.98\hsize}{!}{\includegraphics{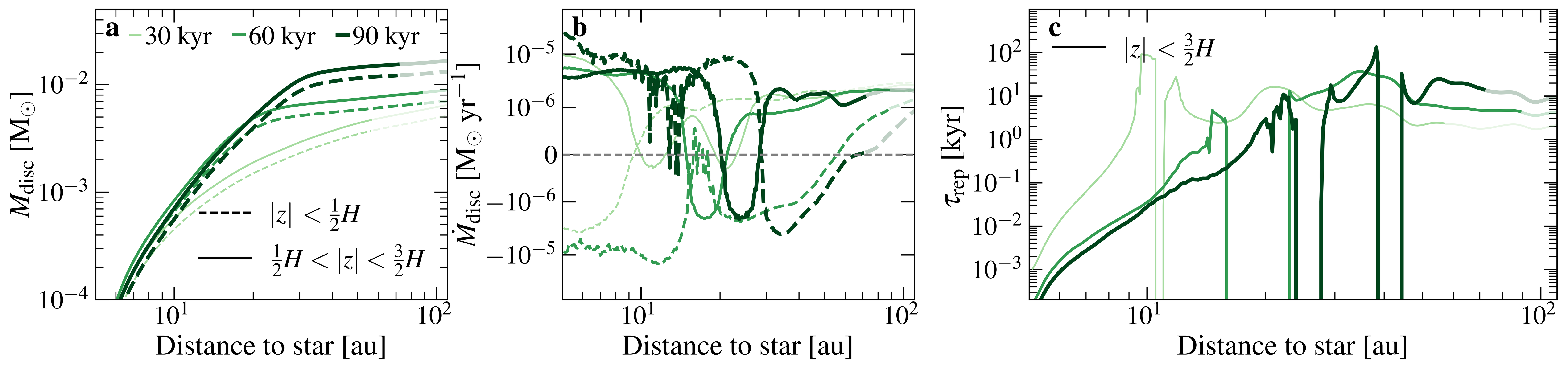}}
    \caption{
    Panel a: Accumulated mass within the disc at a given radius, computed from the fitted surface density. The mass is computed using only 38\,\% and 48\,\% of $\varSigma$, corresponding to the material contained within $|z| < \frac{1}{2}H$ and $\frac{1}{2}H <|z| < \frac{3}{2}H$. 
    Panel b: Integrated mass flux through the lateral surface of a cylinder extending $|z| < \frac{1}{2}H$ and $\frac{1}{2}H <|z| < \frac{3}{2}H$. 
    Panel c: Replenishment time as a function of radius according to the mass and mass flux from panels a and b. 
    }
    \label{fig:disk_masstransport_1D_vertical}
\end{figure*}
As the disc grows with time, the surface density of the gas disc does not change strongly.  
The time evolution of the surface density is shown in panel c of Fig.\,\ref{fig:s180cm5_gendisk}.
Across the outer disc edge, the gas surface density transitions to a shallower decrease with orbital radius, which is in line with a gradual envelope–disc transition zone \citep{das_2026_endtranz}.
For reference, we overplot the classical gas surface density profile for the minimum mass solar nebula (MMSN)
\begin{align}
    \varSigma_{\rm MMSN} = 1700\left( \frac{r}{\rm au} \right)^{-3/2} {\rm g\,cm}^{-2}\,
\end{align}
\citep{hayashi_1981_mmsn}.
Such a power law surface density profile, with $\varSigma \propto r^{-3/2}$, is consistent with an isothermal steady state viscous accretion disc with an accretion rate onto the star of
\begin{align}
    \dot M_{\rm star} = 3 \pi \varSigma \left( \alpha H^2 \varOmega_{\rm kep}\right) \,.
    \label{eq:Mdotdisc}
\end{align}
Here, the term in parentheses is the accretion viscosity, assuming a standard radially-independent accretion parameter $\alpha$ \citep{shakura_Sunyaev}.
Using the gas scale height from our nearly isothermal disc (Eq.\;\ref{eq:Heff}) and 
characteristic values for the disc of system \texttt{180}
, we obtain 
\begin{align}
  \varSigma =& \frac{\sqrt{G}}{3\pi}
   M_{\rm star}^{1/2}\,
   \dot{M}_{\rm star}\,
   \alpha^{-1}\,
   c_s^{-2}\,
   r^{-3/2}\,
   \beta
   \label{eq:surface_density_approximation} \\
   \approx & 60\,{\rm g\,cm}^{-2}
   \left( \frac{M_{\rm star}}{0.6\,{\rm M}_\odot}\right)^{1/2}
   \left( \frac{\dot{M}_{\rm star}}{10^{-5}\,{\rm M}_\odot\,{\rm yr}^{-1}}\right)
   \left( \frac{\alpha}{1}\right)^{-1}\nonumber \\
   &\left(\frac{T}{20\,K}\right)^{-1}
   \left( \frac{r}{10\,{\rm au}}\right)^{-3/2}
   \left(\frac{\beta}{0.1}\right).\nonumber
\end{align}
Here, we used that $(1+\beta^{-1})^{-1}\approx\beta$, when $\beta\ll1$.
While this expression is in approximate agreement with the surface density outside of $10$\,au,
disc dynamics are far from a standard viscous $\alpha$-disc, as illustrated in Section\,\ref{sec:results_flow_and_transport}.
In contrast, the inner magnetically dominated region, where $\beta \propto r^{3/2}$ (Fig.\,\ref{fig:s180cm5_plasma_beta}), is consistent with the 
near-constant surface density with radius. 
This has previously also been seen in global ideal-MHD disc simulations with a decreasing $\beta$ towards the star \citep{suzuki_2014_surface_accretion}, but it may also be induced by overly-efficient parameter-dependent sink-cell accretion (Appendix\,\ref{appendix:sink_cell_prescription}).

Using the gravitational stability criterion Toomre $Q$ \citep{toomre_1964}, we verified that the disc is generally gravitationally stable with surface densities consistently below that of a marginally-stable disc, 
\begin{equation}
    \varSigma_{\rm Q=1} = \frac{\varOmega c_s}{G\pi}.
\end{equation}
In panel c of Fig.\,\ref{fig:s180cm5_gendisk}, we show $\varSigma_{\rm Q=1}$ with a black solid line using the local sound speed, $c_{\rm s}$ and orbital frequency, $\varOmega$ at 90\,kyr.
However, we do note that local regions of the disc may become transiently gravitationally unstable, which we further discuss in Sec.\,\ref{sec:episodic_accretion}.

Finally, 
Figure\,\ref{fig:disk_sizeandmass_s180cm5} shows the radially integrated disc mass, 
\begin{align}
    M_{\rm disc} = 2\pi \int_{r_{\rm in}}^{r_{\rm out}}dr \Sigma r,
    \label{eq:disc_mass}
\end{align}
out to the outer disc edge (green solid line). Here $r_{\rm in}$ and $r_{\rm out}$ are the inner boundary for the radial bins and the fitted disc size presented in Sec.\,\ref{sec:results_disc_size}. In the disc formation stage, the disc rapidly grows to $M_{\rm disc} \approx 10^{-2}$\,M$_\odot$ within $30$\,kyr.
At later times, the disc evolves in tandem with the growth of the host star, maintaining approximately a disc-to-star mass ratio of $\approx$ 4 \%.
During the final $30$\, kyr of simulation time, the increase in disc mass is no longer driven by the growing outer disc edge, but instead through mass loading of the disc, increasing the inner disc surface density (Fig.\,\ref{fig:s180cm5_gendisk}, panel c).

\subsection{Transport of mass through the disc - advection}\label{sec:results_mass_transport}

The concluding step of the analysis examines transport processes confined within the disc. 
We find that mass is efficiently transported inwards and continuously replenished by newly arriving gas accreted by the disc.
We can define a replenishment time as
\begin{align}
    \tau_\mathrm{rep}=\frac{{M_\mathrm{disc}}}{{\dot{M}_{\rm disc}}}\,,
    \label{eq:t_replenishment}
\end{align}
following \cite{kuffmeier_2017}. 
Here $M_\mathrm{disc}(r)$ is the cumulative mass contained within the disc out to a radius $r$ (Fig.\,\ref{fig:disk_masstransport_1D_vertical}, panel a). The mass accretion rate is here defined as 
\begin{equation}
   \dot{M}_{\rm disc}(r)=\int_0^{2\pi}\int_{z_{\rm min}}^{z_{\rm max}}\;d\phi dz\;\rho v_r(r),
\end{equation}
corresponding to the mass flux through the lateral surface of a cylinder at radius $r$. We calculate $\dot{M}_{\rm disc}(r)$ for two surfaces. The midplane lateral surface is set by $|z| < 0.5H$, and we use $0.5H <|z| < 1.5H$ for the disc upper layers. We integrate the gas density, $\rho$ and $v_r$, the radial component of the velocity in cylindrical coordinates, over $d\phi$ and $dz$. Panel b of Figure\,\ref{fig:disk_masstransport_1D_vertical} shows the evolution of the mass accretion rate through the disc as a function of time. 
The upper layers between $0.5H <|z| < 1.5H$ (solid line) are characterised by an inwards mass flux of $\dot{M}\approx10^{-5}$\,M$_\odot$\,yr$^{-1}$ inside of $10$\,au and a decreasing mass flux with increasing radius  
approaching $\dot{M}\approx10^{-6}$ M$_\odot$ yr$^{-1}$ outside of $30$\,au. Additionally, a few narrow regions of outwards mass flux are present, which can, for example, be seen in the interval 20\,au $<r<$ 30\,au at $t=90$\,kyr (Fig.\,\ref{fig:disk_masstransport_1D_vertical}, panel\,b).
Mass transport in the midplane (dashed line) is complex, with initial periods of dominantly outwards transport in the inner disc and more intermittent in/outwards advection in the outer disc (see also Fig.\,\ref{fig:butterfly_60au_s180cm5}).

We find that the replenishment timescale in the outer disc ($r\gtrsim 20$\,au) is approximately $\tau_\mathrm{rep}^{\rm outer}\approx 10$\,kyr, 
well within our simulation time of $t=100$\,kyr (Fig.\,\ref{fig:disk_masstransport_1D_vertical}, panel\,c). 
Near the inner disc edge, replenishment timescales decrease as the enclosed mass $M(r)$ decreases and net ($|z| < 0.5H$) accretion rates $\dot M(r)$ remain high. The timescale of the inner disc is approximately $\tau_\mathrm{rep}^{\rm inner}\approx 0.1$\,kyr at $r\lesssim20$\,au.
The peak in replenishment time around $10$\,au (panel c of Fig.\,\ref{fig:disk_masstransport_1D_vertical}, $t=30$\,kyr) 
is linked to the negative mass flux shown in panel b, indicating outwards flowing gas. 
To conclude, this disc experiences continuous mass loss to the host star and continuous replenishment from infalling material, at a rate that refreshes the full disc mass budget several times over within its simulated lifetime of $t \approx 100$\,kyr.

\section{Diversity between systems}
\label{sec:results_disk_diversity}

Discs form around young stars through continuous filamentary accretion. As indicated by the initial properties of the cores in Table\,\ref{tab:core_overview} 
(Sec.\,\ref{sec:resutls_initial_condition}), the nine systems have slightly different initial conditions and therefore the evolution of the systems varies. 
In the following subsections, we first describe the evolution of the accretion rates around these nine systems, followed by a characterisation of streamer infall events, and conclude with a discussion of the disc radius and mass evolution.

\subsection{Accretion rates onto the host star}\label{sec:results_stellar_accretion}

\begin{figure}
    \centering
    \resizebox{\hsize}{!}{\includegraphics{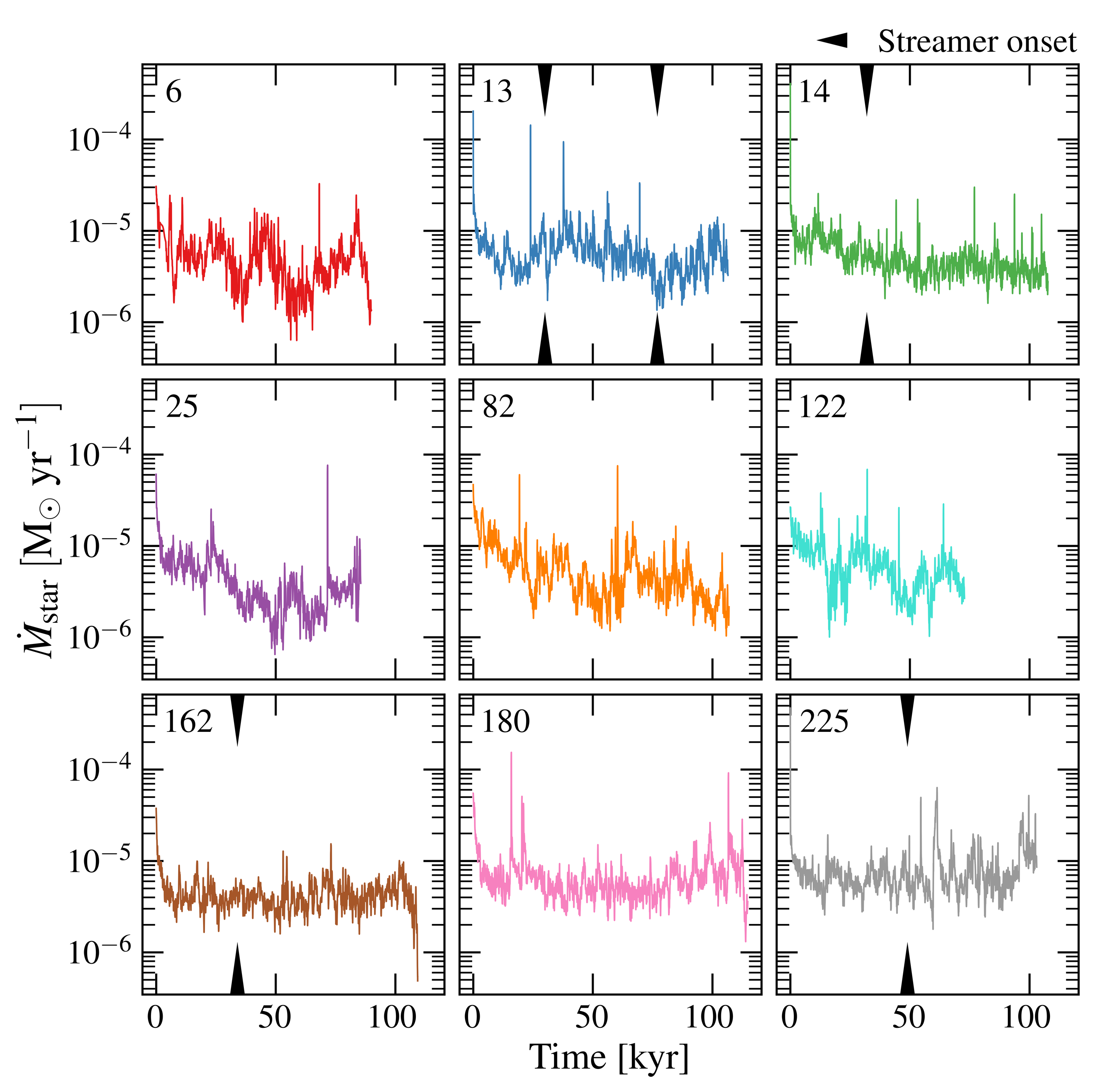}}
    \caption{Computed accretion rate for the nine systems as a function of time. The star data has a time resolution of  10 years, making it the data with the highest cadence. The triangles signify the time of the snapshots seen in Fig.\,\ref{fig:combined_streamer_plot}.}
    \label{fig:only_sink_accretion}
\end{figure}
\begin{figure}
    \centering
    \resizebox{\hsize}{!}{\includegraphics{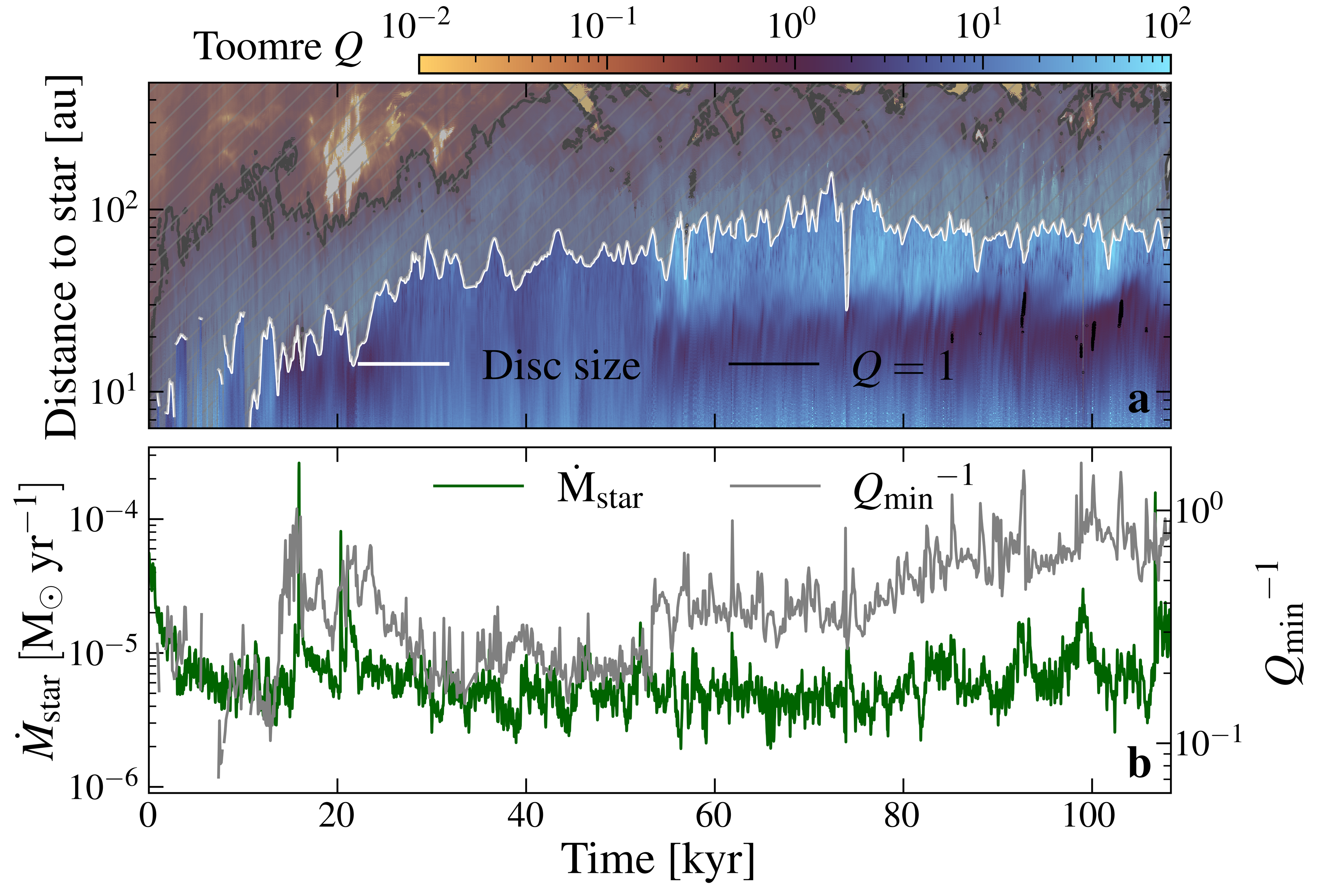}}
    \caption{
    Panel a: 
    Time evolution of the Toomre $Q$ parameter as a function of radius for system \texttt{180}.
    The grey-hatched region marks the region where the Toomre $Q$ parameter has no clear physical meaning, corresponding to periods when no disc is present or locations outside of the outer disc edge marked with a white line. 
    The disc remains globally gravitationally stable at all times. However, during the disc evolution stage ($t\gtrsim50$\,kyr), a persistent darker band of low $Q$ appears near $\sim$\,20\,au, within which the $Q=1$ contour indicates the disc becomes only marginally stable ($Q\approx1$).
    Panel b: Stellar accretion rate (green, left y-axis) as a function of time for system \texttt{180} and the reciprocal minimum Toomre $Q$ value (grey, right y-axis), so that instability epochs appear as peaks alongside the accretion bursts.
    Episodic accretion events are highly correlated with periods of marginal instability ($Q^{-1}\gtrsim1$).} 
    \label{fig:ToomreQ_timeevo_s180cm5}
\end{figure}

\begin{figure*}
    \centering
    \resizebox{\hsize}{!}{\includegraphics{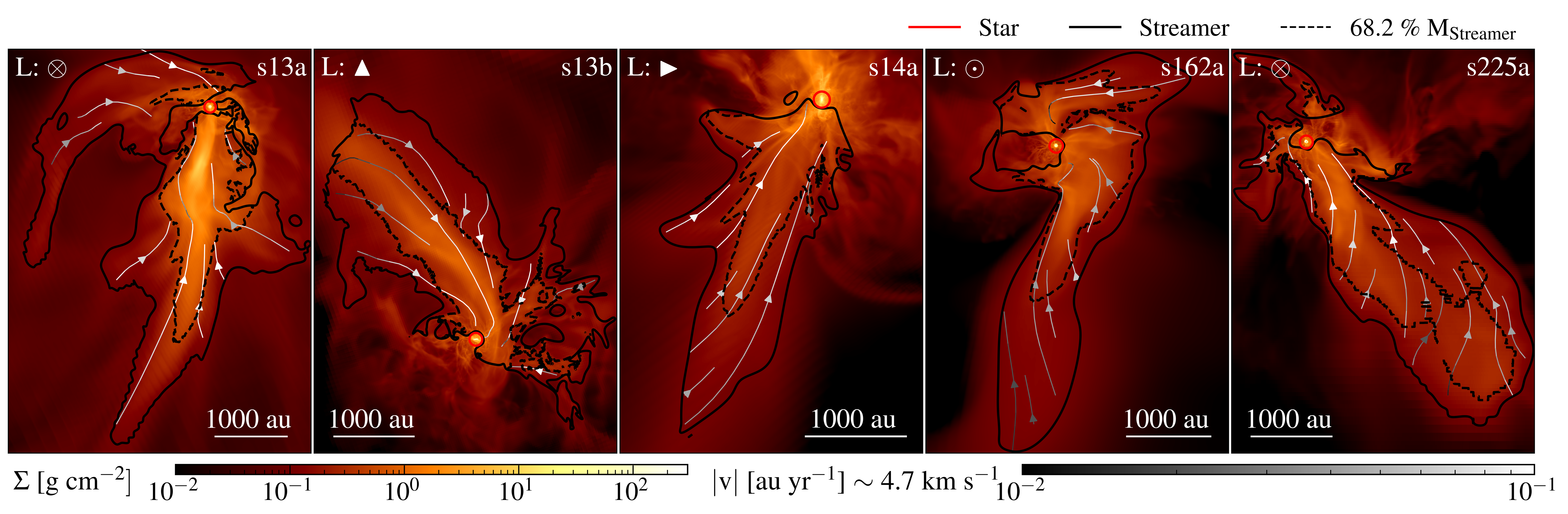}}
    \caption{Five streamers in four different systems. The integrated depth for all plots is 2000 au. The colours display the integrated density, and the streamlines are the projected velocity in au yr$^{-1}$. The solid black line marks the streamer boundary, while the dashed black line indicates the region, containing 68.2 \% of the streamer mass. The red solid line shows the star-disc region. Each streamer is labelled in the top-right corner; for example, \texttt{s13a} refers to a streamer in system \texttt{13}, with the letter denoting the specific streamer. The orientations of the plots are made with respect to the angular momentum vector within 50 au of the star, and can be seen in the top left corner of each panel. The plots are oriented as follows: \texttt{s13a} is face-on, rotating clockwise. \texttt{s13b} is edge-on with the spin axis in the positive y-direction. \texttt{s14a} is edge-on with the spin axis pointing in the positive $x$-direction. \texttt{s162a} is face-on, rotating counterclockwise. Finally, \texttt{s225a} is face-on, rotating clockwise.}
    \label{fig:combined_streamer_plot}
\end{figure*}

Typically, our stars initially accrete at high rates, $\dot{M}_{\rm star}$$\sim$$10^{-4}$\,M$_\odot$\,yr$^{-1}$, whereafter accretion rates decline to around $10^{-5}$ M$_\odot$\,yr$^{-1}$ within the first $5$\,kyr of evolution (Fig.\,\ref{fig:only_sink_accretion}). 
Around $t\approx10^5$\,yr, stellar accretion rates generally have decreased to $5\times 10^{-6}$ M$_\odot$ yr$^{-1}$.
We note that \citet{kuffmeier_2017} report some systems showing more rapidly decreasing stellar accretion rates, approaching $10^{-6}$\,M$_\odot$\,yr$^{-1}$ after already $50$\,kyr of evolution, presumably because their giant molecular cloud environment is less dense ($1.6\,{\rm M}_\odot\,{\rm pc}^{-3}$), compared to our model ($47\,{\rm M}_\odot\,{\rm pc}^{-3}$). Interestingly, our ideal MHD results differ strongly from the non-ideal simulations in \citet{lebreuilly_2024a} where the stellar accretion rates increase by up to two orders of magnitude during the simulation period of $\sim$10$^{5}$\,yr.

The stellar accretion rates obtained in this study are in broad agreement with observationally determined accretion rates onto young stars. 
Observations of deeply-embedded Class 0 stars, with ages $\lesssim$$10^5$\,yr \citep{evans_2009_spitzer_class0_age}, point to accretion rates in the range between 
$\dot{M}_{\rm star}$$\approx$$5\times 10^{-7}$ M$_\odot$ yr$^{-1}$
and 
$\dot{M}_{\rm star}$$\approx$$5\times 10^{-5}$ M$_\odot$ yr$^{-1}$
\citep{watson_2016_class0_accretion_rates}, making the here modelled accretion rates broadly consistent with the upper segment of these estimates. 
Older Class I sources accrete at rates around $\dot{M}_{\rm star}$$\sim$$10^{-6}$\,M$_\odot$\,yr$^{-1}$ \citep{fiorellino_2023_class1_accretion_rates}, after which accretion rates further decline in class II discs \citep{manara_2012}.

\subsection{Episodic accretion}\label{sec:episodic_accretion}

We find that all systems, except system \texttt{162}, show short accretion peaks where the accretion rates suddenly increase by one to two orders of magnitude on a short timescale of approximately $\sim$\,$100$\,yr. In particular, systems \texttt{13}, \texttt{14} and \texttt{122} exhibit prominent accretion bursts with strength of $\dot{M}_{\rm star}$$\sim$$10^{-4}$\,M$_\odot$\,yr$^{-1}$, recurring with an approximate minimum timescale of $10$\,kyr. 
These findings align with observations of episodic accretion for young stellar objects where accretion rates increase by up to three orders of magnitude with a characteristic lifetime of $50$-$100$\,yr, recurring every $5$-$50$\,kyr \citep{audard_2014_episodic_accretion}.
The origin of these outbursts in our simulations is correlated with short-lived local regions of the disc that are weakly gravitationally unstable and drive temporary spiral waves that transport angular momentum and drive increased accretion rates
\citep[][Sec.\,\ref{sec:discfrag}]{kuffmeier_2018}. 
Panel a and b of Figure\,\ref{fig:ToomreQ_timeevo_s180cm5}
illustrate that
regions of $Q_{\rm min}\approx1$ are strongly correlated with accretion peaks (green line, panel b). 
In the calculation of Toomre $Q$ we used the fitted surface density (see Appendix\,\ref{sec:appendix_disc_fitting_procedure}) and measured $c_{\rm s}$ and $\varOmega$ as mass-weighted averages within 3 scale heights.
In contrast, the massive streamers we have identified (see Sec.\,\ref{sec:results_streamers}), do not always appear strongly correlated with accretion spikes, but instead align with periods of strong disc truncation, as we further discuss below.
Finally, we note that some of our systems show long-period variations on timescales of $10$--$20$\,kyr in the accretion rate by up to a factor of 5, e.g. systems \texttt{6}, \texttt{82} and \texttt{122}. In contrast, systems \texttt{162}, \texttt{180}, and \texttt{225} display stable accretion periods with no to few signs of episodic accretion or any clear long-period oscillations. These three systems originate from the least mass-loaded cores, having the lowest initial core masses of our sample (M$_\mathrm{core}\leq$\,1.68\,M$_\odot$,
Table\,\ref{tab:core_overview}).

\subsection{Characterisation of streamer events}\label{sec:results_streamers}

\begin{table}
    \centering
    \caption{
    Properties for the streamers shown in Figure\,\ref{fig:combined_streamer_plot}.
    The displayed quantities are 
    onset of the streamer $t_{\rm str}$, 
    total mass of the streamer excluding the region 
    within $100$\,au of the star
    $M_{\rm str}$, 
    average streamer column density $\langle\varSigma\rangle_\mathrm{str}$, 
    environment column density $\langle\varSigma\rangle_\mathrm{env}$, 
    streamer length l$_{\rm tot}$, and average velocity within the streamer
    $\langle v\rangle$. 
    The last two rows contain analysis 
    considering the central 68.2 \% of the mass in each streamer. 
    The average streamer accretion, $\langle\dot{M}\rangle_\mathrm{str}$, is the mass over duration, assuming all material effectively lands in the system.
    Here, $\Delta t_{\rm str}$ is a proxy for the streamer duration,  
    and 
    $\langle\dot{M}\rangle_\mathrm{str} = 0.682M_{\rm str}/ \Delta t_{\rm str}$ the approximate streamer accretion rate.
    }\label{tab:streamer_overview} 
    \setlength{\tabcolsep}{4pt} %
    \begin{tabular}{llccccc}
    \hline\hline
      \multirow{2}{*}{Quantity\hspace{-4cm}} & 
      \multirow{2}{*}{} & 
      \multirow{2}{*}{\texttt{s13a}} & 
      \multirow{2}{*}{\texttt{s13b}} & 
      \multirow{2}{*}{\texttt{s14a}} &
      \multirow{2}{*}{\texttt{s162a}} &
      \multirow{2}{*}{\texttt{s225a}} \\
      &
      &
      &
      &
      \\ \hline           
        $t_{\rm str}$                       & [kyr]                                 & 30    & 77    &  32    &  34   & 49  \\
        $M_\mathrm{str}$                       & [M$_\odot$]                           & 0.39 & 0.18 & 0.088 &  0.13  & 0.17 \\
        $\langle\varSigma\rangle_\mathrm{str}$ & 10$^{-2}$[g cm$^{-2}$]     & 43   & 26   &  26   &  24  &  25 \\
        $\langle\varSigma\rangle_\mathrm{env}$     & 10$^{-2}$[g cm$^{-2}$] & 7.4  & 6.6  &  6.6  &  4.1 &  4.2 \\
        $l_{\rm tot}$                     & 10$^3$[au]                                  & 4.9  & 3.4  &  3.7  &  3.7 & 4.1   \\
        $\langle v\rangle$& [au yr$^{-1}$]                        & 0.06  & 0.07  &  0.07  &  0.05 &  0.06 \\  \hline
         \multicolumn{7}{l}{\raisebox{0pt}[2ex][1ex]{Analysis limited to 68.2 \% of mass}} \\ \hline
         $\Delta t_{\rm str}$                       & [kyr]                                  & 8.5   &  10  &  6.6  &  6.5   & 11   \\
         $\langle{\dot{M}}\rangle_\mathrm{str}$ &10$^{-5}$[M$_\odot$ yr$^{-1}$]  & 3.2   & 1.2   &  0.92  &  1.4   & 1.0 \\
    \hline\hline
  \end{tabular}
\end{table}

A survey of the evolution of the nine systems indicates at least
five massive streamers.  
Here, we define a streamer as 
(i) an over-dense region with a column density exceeding $0.1$\,g cm$^{-2}$ along a $2000$\,au integration depth and 
(ii) an infalling motion towards the star-disc system, requiring an angle less than $\pi/2$ between the mass-averaged velocity vector and the star-position vector in the projected plane.
In this calculation, we omit the innermost region close to the star $<100$\,au.
The streamers identified in this way are shown in Figure\,\ref{fig:combined_streamer_plot} for systems \texttt{13}, \texttt{14}, \texttt{162} and \texttt{225}.
The shown streamers are several $1000$\,au long similar to observational studies reporting the presence of streamer-like structures, similar in size and morphology, extending from several thousand au \citep{alves_2020, pineda_2020_huge_class0_streamer_mass, flores_2023_obs_streamers}. We calculate the streamer length, $l_{\rm tot}$, as the distance from the star to the farthest contained point (Table\,\ref{tab:streamer_overview}).
We find that the average streamer column density is 0.28\,g\,cm$^{-2}$. 
This can be compared to the average background surface density of the environment, $\langle\varSigma\rangle_\mathrm{env}$ beyond 500\,au from the central star (Table\,\ref{tab:streamer_overview}). Generally, the column density contrast between streamers and their surroundings ranges between a factor 4 to 6.

These streamer infall events last for around  $\Delta t_{\rm str} \approx 10$\,kyr (Table\,\ref{tab:streamer_overview}).
We determined the duration $\Delta t_{\rm str}$ as the average free-fall time of the streamer
gas parcels, assuming they freely accelerate towards the star, which agrees well with visual inspection of their presence in simulation snapshots (see also Fig.\,\ref{fig:azimuthal_3avg_dmdtheta_150au_s13cm5_fix}).

In terms of mass, the streamers range from $M_{\rm str}$\,$=$\,$0.088$\,M$_\odot$ to $M_{\rm str}\,=\,0.39$\,M$_\odot$. 
Our measured streamer masses lie within the estimate for the Class 0 object IRAS 03292+3039 \citep[0.1-1\,M$_\odot$,][]{pineda_2020_huge_class0_streamer_mass}.
The more massive structures generally drive higher accretion rates.
The highest accretion rate is associated with streamer \texttt{s13a}, delivering $\dot M_{\rm str}=3.2$\,$\times$\, $10^{-5}$\,M$_\odot$\,yr$^{-1}$, and the lowest with streamer \texttt{s14a}, $\dot M_{\rm str}=0.92$\,$\times$\,$10^{-5}$\,M$_\odot$\,yr$^{-1}$ (Table\,\ref{tab:streamer_overview}). 

In general, the mass delivery rate of the streamers is of the same order of magnitude as the stellar accretion rate in these systems of $\dot M_{\rm star}\sim 10^{-5}$\,M$_\odot$\,yr$^{-1}$. 
Such a similarity between streamer and stellar accretion rate has also been observed in the VLA1623 B system, although for a lower accretion rate regime, with 
$\dot M_{\rm str} =$3-5\,$\times$\,10$^{-7}$\,M$_\odot$\,yr$^{-1}$ and 
$\dot M_{\rm star}=$0.6-3\,$\times$\,10$^{-7}$\,M$_\odot$\,yr$^{-1}$ \citep{codella_2024_streamer_mass}.

\subsection{Impact of streamer infall onto discs}
\begin{figure}
    \centering
    \resizebox{\hsize}{!}{\includegraphics{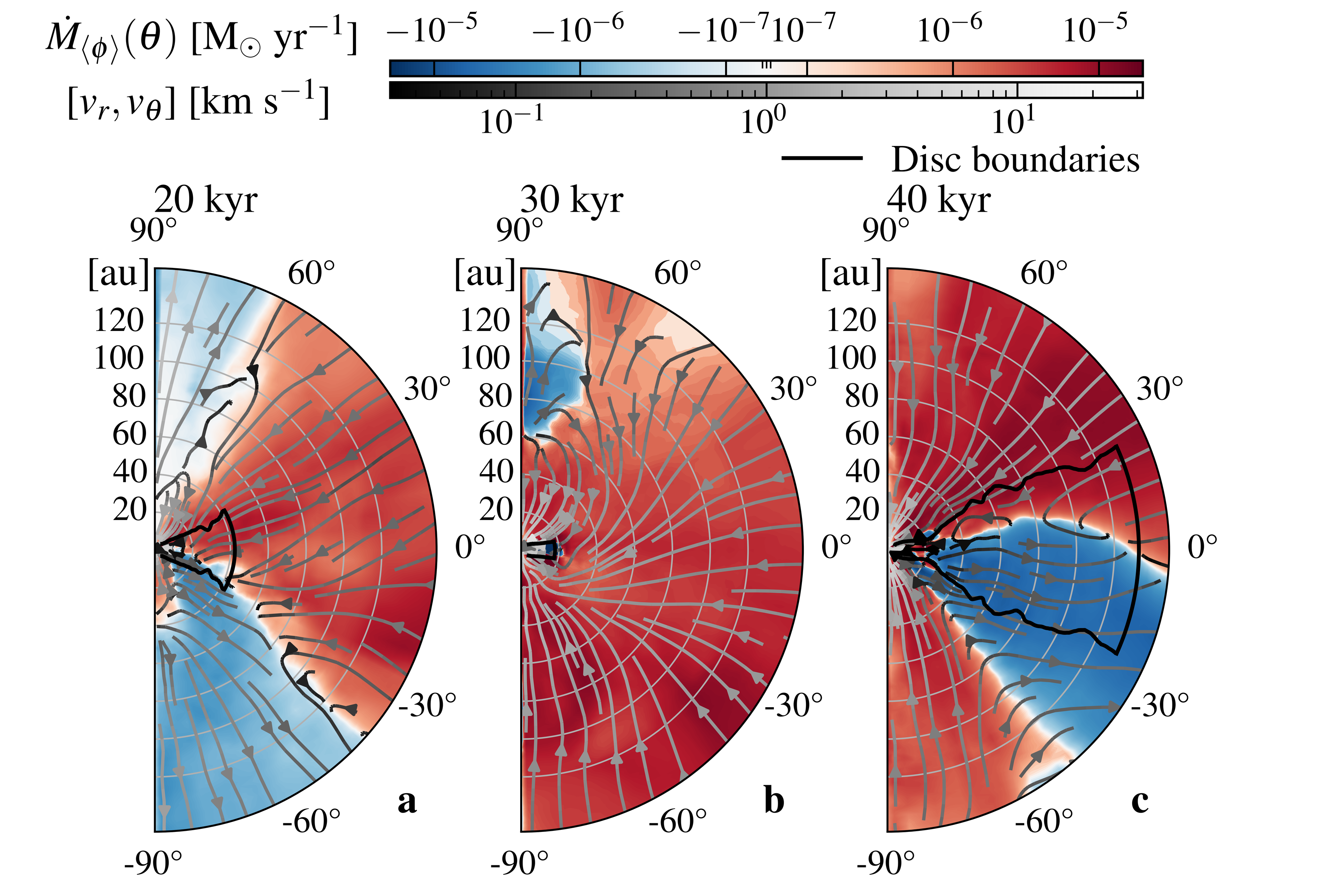}}
    \caption{
    %
    Snapshots of mass accretion as a function of polar angle at different times ($t=20,30,40$\,kyr) for system \texttt{13}. 
    The flow lines show the mass-averaged poloidal field flow. 
    The panels are azimuthally and temporally averaged over a dynamical timescale of $1$\,kyr.
    The disc boundaries, as shown in Sec.\,\ref{sec:results_surfac_density}, are plotted with a black contour.
    Panel a ($t=20$\,kyr) shows prominent outflows along the polar axis ($|\theta|>60^\circ$).
    Panel b ($t=30$\,kyr) corresponds to the onset of the streamer \texttt{s13a} reaching the disc.
    Finally, panel c ($t=40$\,kyr) illustrates the disc and envelope state when streamer infall comes to an end.
    \label{fig:azimuthal_3avg_dmdtheta_150au_s13cm5_fix}
    }
\end{figure}

The most massive streamer we identified, \texttt{s13a}, deposits a total mass of $0.4$\,M$_\odot$, disrupting the existing disc morphology.
First, the disc gets truncated, but this is followed by a period of about $10$\,kyr where the disc rapidly grows in radius.
This is shown in Figure\,\ref{fig:azimuthal_3avg_dmdtheta_150au_s13cm5_fix}, displaying the poloidal mass-averaged velocity field, along with mass flux per unit polar angle (Eq.\,\ref{eq:massflux_per_unit_polarangle}). 
Investigating this event in more detail shows that strong polar outflows are present $10$\,kyr before the onset of the streamer (panel a of Fig.\,\ref{fig:azimuthal_3avg_dmdtheta_150au_s13cm5_fix}).
Later, as the streamer reaches the system (panel b, Fig.\,\ref{fig:azimuthal_3avg_dmdtheta_150au_s13cm5_fix}), the system 
gets infall-dominated, and the disc size decreases to approximately $r_{\rm disc}=10$\,au.
Especially, the region below the midplane, $\theta<0^\circ$, is dominated by a large flux of material,  
which leads to a re-alignment of the disc by about $90^\circ$. When the infall of this massive streamer has come to an end (panel c, Fig.\,\ref{fig:azimuthal_3avg_dmdtheta_150au_s13cm5_fix}), a significant outwards mass flux in the midplane $\theta\lesssim10^\circ$ increases the outer disc radius to approximately $r_{\rm disc}=130$\,au, while above $\theta\gtrsim10^\circ$, material is still infalling towards the system with a mass flux of $10^{-5}$ M$_\odot$\,yr$^{-1}$. 
The rapid disc destruction by the streamer-disc interaction motivates a new classification: {\it  destroyer-class} streamer (with $M_{\rm str} >  M_{\rm disc}$).

It is not straightforward to identify why some systems undergo prominent streamer infall events, while others do not.
For example, system \texttt{13} exhibits two prominent streamers and a high degree of fluctuation in both stellar accretion (Fig.\,\ref{fig:only_sink_accretion}) and disc size (presented in Sec.\,\ref{sec:all_disc_sizes}).
This can be placed in comparison with the much calmer system \texttt{180} (Sec.\,\ref{sec:results_disc_size}). A possible origin could be the difference in the magnetisation of the cores from which they form:  core \texttt{13} has a stronger mean magnetic field, $\langle B\rangle=$\SI{51}{\micro G}, and shows a larger field dispersion, $\sigma_B=$\SI{85}{\micro G}, while \texttt{180} contains a weaker mean field, $\langle B\rangle=$\SI{31}{\micro G} and less dispersion, $\sigma_B$=\SI{38}{\micro G}. 
In fact, we note that system \texttt{180} is exceptional in having a low magnetic field strength $\langle B\rangle$, a low mass budget $M$, and low rotational energy $E_\mathrm{rot}$ (1$\sigma$ below the mean values of all cores listed in Table\,\ref{tab:core_overview}).
These characteristics suggest that system \texttt{180}, originating from a low-mass, weakly magnetised, and low-energy core, shows
a comparatively calmer dynamical evolution, with reduced turbulent activity relative to systems formed from more massive and strongly magnetised cores.

In summary, while all systems share broadly similar stellar masses and accretion rates, their individual evolutionary paths diverge, with some showing episodic bursts or long-period oscillations, and others evolving more steadily. 
This preliminary analysis of the data has shown that, for at least some of the systems, the mass transport happens anisotropically in narrow streams.
The streamers are transient events and do not persist throughout the entire evolution. 
The accretion rates for the systems without massive streamers, like system \texttt{180}, are
still in the order of  10$^{-5}$ M$_\odot$ yr$^{-1}$. This implies that the systems experience a constant "background" accretion, and in the case of an infalling streamer, it adds to the mass flux, in the same order of magnitude as the accretion background. 
A more comprehensive analysis of streamer counts across the entire dataset, along with a detailed characterisation, is planned for future work.

\subsection{Evolution of the disc size}\label{sec:all_disc_sizes}
\begin{figure}
    \centering
    \resizebox{\hsize}{!}{\includegraphics{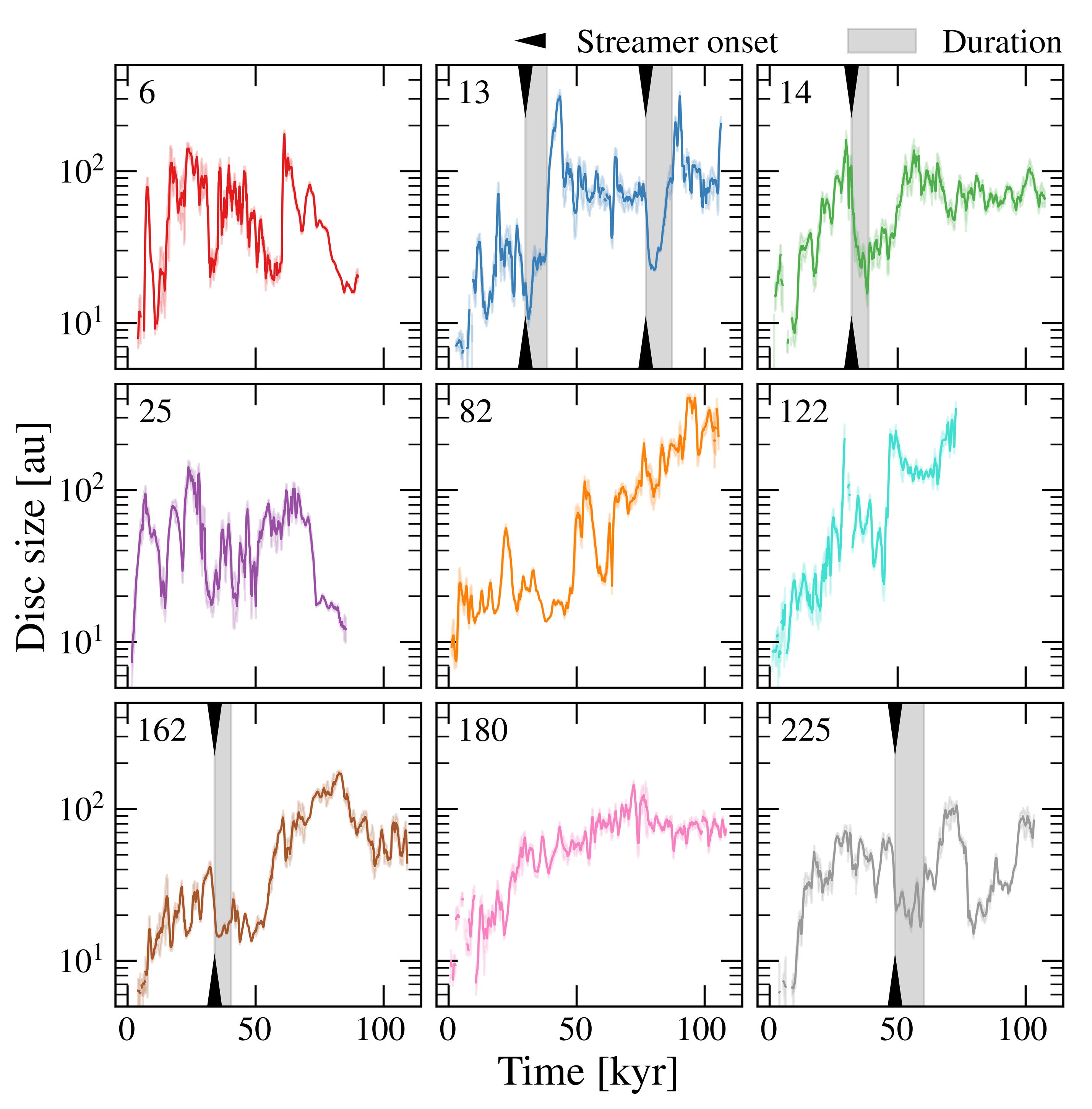}}
    \caption{
    Evolution of disc size as a function of time, for all considered systems.
    The one-$\sigma$ uncertainty on the outer disc radii, obtained after $1$\,kyr time-averaging, is shown with lower opacity.
    The streamer onset times are displayed with triangles in the top and bottom, while the grey region marks the duration of the infall (Table\,\ref{tab:streamer_overview}).
    In general, discs form after roughly 15\,kyr and can grow to sizes of up to several $\sim100$\,au by the end of the integration period ($\sim10^5$\,yr). Streamer accretion events temporarily shrink the discs, after which they recover to similar sizes or even grow to larger radii.
    }
    \label{fig:all_disc_sizes}
\end{figure}

In general, we can identify that all discs undergo a disc formation period, where the disc grows to several tens of au within the first roughly $20$\,kyr.
Figure\,\ref{fig:all_disc_sizes} shows this general trend.
The diverse subsequent evolution is driven by the continuous filamentary accretion and major streamer infall events, the latter marked with black 
triangles in the top and bottom. Disc radii are calculated in the same fashion as described in Sec.\,\ref{sec:results_disc_size}.
In particular, systems \texttt{13}, \texttt{14}, \texttt{162}, and \texttt{225} illustrate how disc sizes often decrease, by typically as much as $20$ to $60$\,au, corresponding to an average reduction by approximately 65\,\%.
The deposition of streamer material corresponds to a major re-arrangement of the angular momentum budget of the disc.
After the truncation and the end of the streamer infall (grey bands), disc sizes typically expand from their prior size on a $10$\,kyr timescale. 

Broadly speaking, the longer-term evolution of disc radii is regulated by the cloud environment. 
More quiescent cores, like \texttt{162} and \texttt{180}, result in efficient disc growth to a relatively stable disc radius of $r_{\rm disc} \approx 100$\,au after $50$\,kyr.
In contrast, systems \texttt{6}, \texttt{13}, \texttt{14}, \texttt{82}, and \texttt{122} that form in more dynamic environments 
show more fluctuating disc radii as a result. This variability can also be seen in the evolution of their stellar accretion rates with time, showing episodic bursts and long-period oscillations (Fig.\,\ref{fig:only_sink_accretion}).
Interestingly, \texttt{82} and \texttt{122} do not yet approach a quasi-steady state during their $10^5$\,yr of evolution. Instead, we find still growing discs, reaching radii of several hundred au wide, around their central sub-solar stars ($M_{\rm final} \approx 0.5\,M_\odot$). 

The gas disc radii measured here are broadly consistent with those inferred around young stars. Observations find dust disc radii in the $50$ to $100$\,au range around isolated Class 0 stars with masses above $0.5$\,M$_\odot$ \citep{yen_2024_edisk_sizes}. 
However, modelling of synthetic observations show the effective gas radius for stars in that mass range is about a factor 1.5 times larger than the mm continuum dust radius \citep{aso_2020_synthetic_observations}. 
For example, the well-characterised Class 0 Protostar L1527 has a dust radius of about $85$\,au, but a gas radius of $\approx$\,$110$\,au as determined from the CO rotation curve \citep{vant_Hoff_1_observation_of_gas_and_dust_disc}.

We find that disc truncation is a natural outcome of destroyer-class streamer-disc interaction, which may be linked to stellar accretion outbursts as well.
Observations of stars undergoing FU Orionis-like variable accretion, show that they typically are accompanied by relatively compact discs, 
with dust radii between 16 and 69\,au, which are 1.5--4.7 times smaller than typical discs around Class I/II objects \citep{kospal_2021_FU_orionis_ALMA}. Possibly, the variability in stellar accretion rates and reductions of the outer disc radii hint at past streamer infall events.

\subsection{Evolution of the disc mass}\label{sec:all_disc_masses}

\begin{figure}
    \centering
    \resizebox{\hsize}{!}{\includegraphics{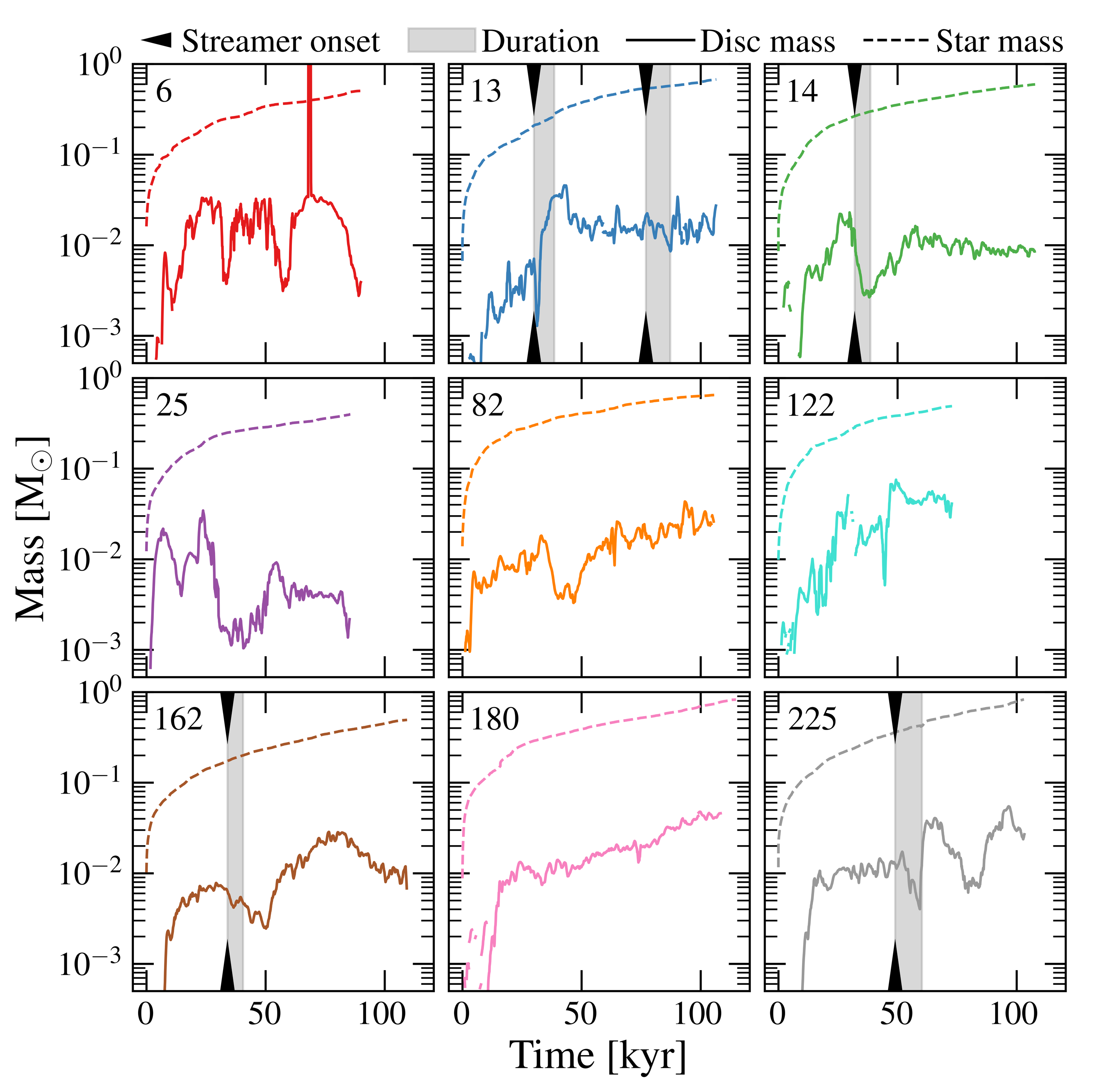}}
    \caption{
    Evolution of disc and star mass as a function of time for all systems. 
    The disc masses shown here are the accumulated mass enclosed within $z\pm3H$ integrated to the outer disc edge (shown in Fig.\,\ref{fig:all_disc_sizes}). The disc masses show a similar evolution to the disc sizes. However, they do not lose equivalent mass to the reduction in size.
    }
    \label{fig:all_disc_masses}
\end{figure}

Disc masses are highly correlated with disc sizes, as discs primarily grow in mass through the addition of mass to the outer disc edge (Sec.\,\ref{sec:results_disc_size}). 
Infall events that trigger a disc size reduction also reduce the disc in mass (Fig.\,\ref{fig:all_disc_masses}, solid line), typically by about 40\,\%.
Star masses increase over time and reach final masses between 0.39\,M$_\odot$ and 0.83\,M$_\odot$ within the integrated time. This is displayed in Figure\,\ref{fig:all_disc_masses} (dashed line). The disc-to-star mass ratio is consistently below 10\,\% for all systems.
Interestingly, in \cite{kuffmeier_2017} some systems show a steady decrease in disc mass 
with time, such that after $50$\,kyr of evolution disc masses are around $M_{\rm disc}\lesssim\,10^{-3}\,M_\odot$, corresponding to approximately disc-to-star mass ratios of $M_{\rm disc}/M_{\rm star}\lesssim\,10^{-2}$.
These rapidly evolving discs occur typically around lower mass stars ($M<0.4$\,M$_\odot$), below the stellar mass cut of the sample presented here.
Finally, we briefly note that the apparent spike visible at 68 kyr for system \texttt{6} in Figure\,\ref{fig:all_disc_masses} is due to a temporary failure of the disc mass fitting routine 
during an epoch where the vertical density profile deviates strongly from hydrostatic balance (Eq.\,\ref{eq:vertical_structure}).



\section{Comparison to previous work}\label{sec:discussion_previous_work}

\subsection{Non-ideal MHD}
Disc formation proceeds more rapidly in simulations including ambipolar diffusion which reduces magnetic braking 
because of magnetic flux diffusion and physical dissipation \citep{Li_2014_magnetic_reconnection,tomida_2015_nonideal_limits_magneticbraking, masson_2016_discsizes}. 
Recently, \citet{mayer_2025_largescale_niMHD_structures} presented high-resolution zoom-in simulations, using the \textsc{arepo} code, resolving the collapse down to the second Larson core, with minimum cell sizes of $\sim$\,10$^{-4}$\,au. 
In the ideal MHD limit, \citet{mayer_2025_largescale_niMHD_structures} report a lack of discs forming within their simulation time of $10^4$\,yr, as opposed to simulations that include ambipolar diffusion. This is broadly consistent with our finding, where we typically see that discs need an initial growth time of several times $10^4$\,yr before emerging (Fig.\,\ref{fig:all_disc_sizes}). 
This therefore aligns with the general finding that discs also form under ideal MHD conditions when turbulence is accounted for  \citep{santos-lima_2012_turbulence_and_Bfield, seifried_2012_turbulence_and_Bfield, joos_2013_turbulence_and_Bfield, wurster_2019_ni_i_disc_sizes}.

Similarly, \citet{lebreuilly_2024a} compared disc formation with ideal MHD and MHD including ambipolar diffusion, in zoom-in simulations including flux-limited radiative transfer using the \textsc{ramses} code.
Discs also form under ideal MHD conditions, but are more strongly magnetised with global plasma-$\beta$ values around unity, compared to their non-ideal simulations with an order of magnitude higher plasma-$\beta$. They also found that these weaker magnetic fields lead to an increase in disc sizes by a factor of two relative to strongly magnetised discs \citep{lebreuilly_2024a}.

Given these previous results, it is therefore important to stress that the results presented in this work are done in the ideal MHD limit. 
This is a crude approximation during the initial stages of disc growth on small scales $\lesssim 10$\,au \citep{masson_2016_discsizes}.
However, the ideal MHD limit is more appropriate at wider orbital distances where cosmic rays are an efficient source of ionisation. 
Figure\,\ref{fig:s180cm5_gendisk}, panel c, illustrates that disc surface densities do not typically exceed $\Sigma_{\rm g}=100$\,g\,cm$^{-2}$ outside of $r=10$\,au during formation, allowing cosmic ray ionisation down to the midplane of these discs \citep{umebayashi_1981_origin_ionisationrate}. 
This is because the disc primarily grows through continued radial accretion of material, rather than viscous relaxation of an initial compact high-surface-density disc \citep{lynden_bell}.
Moreover, recent observations hint that effective ionisation rates are high near young stellar objects \citep[up to $\zeta=10^{-16}$\,s$^{-1}$ within $r=10^4$\,au, ][]{pineda_2024_cosmicrays}, a factor ten higher compared to a standard value for prestellar cores \citep[$\zeta=10^{-17}$\,s$^{-1}$,][]{spitzer_1968_CR_rate,caselli_1998_CR_rate_observation}. 
Two recent observations of Class 0 protostars B335 ($M_{\rm B335}=0.1-0.4\,{\rm M}_\odot$, protostar) and L1157 
\citep[$M_{\rm L1157}\sim2\,{\rm M}_\odot$, envelope,][]{gueth_2003_L1157_envelopemass}
show values of $\zeta=10^{-14}$\,s$^{-1}$ and $\zeta=$10$^{-13}$\,s$^{-1}$  \citep{cabedo_2023_ionisationrate, schwarz_2026_high_ionisation}.
However, closer to the host star, disc midplanes may be partly shielded from ionising radiation by stellar winds \citep{cleeves_2013_discionisationfromcosmicrays}, although it is unclear if this mechanism is present when outflows are disc-dominated and instead guided along the disc magnetic field \citep{fujii_2022_shieldingofcosmicryas_magneticfields}.

\subsection{Envelope and disc structure}
In simulations of an isolated non-ideal MHD collapse of a single core, the envelope flattens along the magnetic field into a pressure-supported non-Keplerian pseudo-disc, which then feeds the protostar-disc system with low angular momentum material \citep{xu_kunz_2021a_singlecore, xu_kunz_2021b_singlecore}. 
In our simulations, the envelope is instead shaped by turbulent infall, which makes it difficult to identify the emergence of a non-Keplerian initial disc \citep{mayer_2025_largescale_niMHD_structures}. 
In such a turbulent environment, filamentary infalling sheetlets can represent a disrupted form of the classical pseudo-disc \citep{tu_2024a, tu_2024b}. We indeed identify filamentary infalling structures similar in column density ($\sim$10$^{1}$\,g\,cm$^{-2}$) and spatial scale ($\sim$10$^{2}$\,au), see Fig.\,\ref{fig:osyris_rho_7panel_s180cm5}, that are comparable to the filaments in their least diffusive model \citep[][their Fig.\,2, M1.0AD0.1]{tu_2024a}, which is closest to the ideal MHD limit considered here.

The strong filamentary inflow and epochs of massive streamer infall, that are inherited from large supersonic turbulence, appear to largely suppress strong outflows \citep{Padoan_2020_unboundaccreting_material}. 
However, our resolution is not yet sufficient to resolve collimated jet formation (Appendix.\,\ref{appendix:numerical_resolution}), which could be an important source of feedback on the surrounding ISM \citep{zanni_2007_jet_efficiency, tu_2026_highvelocityjets}. 
Therefore, an important avenue for future work is to quantify these outflows better,  which could then improve outflow prescriptions for our lower-resolution ($25$\,au) parental star-formation runs that currently use fixed outflow-loss fractions of $50$\% \citep[][their Sec.\,2]{haugbølle_2018}.

An interesting aspect of the disc evolution phase of system \texttt{180} is the surface-layer accretion and toroidal field reversal previously seen in other ideal MHD isolated disc simulations \citep{suzuki_2014_surface_accretion, zhu_stone_2018}. Surface accretion is also seen in non-ideal disc simulations \citep{suriano_2017_Bfield_transport_layer,lee_2021_surfacelayer}. However, in the non-ideal regime the inner disc is demagnetised with plasma-$\beta$ values increasing towards the host star and a more quiescent midplane with low turbulence.

\subsection{Disc fragmentation}
\label{sec:discfrag}
In our simulations discs remain globally gravitationally stable throughout the simulated evolution, with Toomre $Q$ values consistently above unity 
except for localised, transient episodes, likely associated with enhanced infall (Sec.\,\ref{sec:episodic_accretion}, Fig.\,\ref{fig:ToomreQ_timeevo_s180cm5}). 
The Toomre $Q$ parameter for system \texttt{180} can be seen in Figure\,\ref{fig:ToomreQ_timeevo_s180cm5}, and shows a stable disc within the outer disc radius through the full evolution. However, at times $t\approx20$\,kyr and $t\gtrsim 50$\,kyr we see $Q\approx1$ signifying a marginally unstable disc.
This contrasts with disc simulations without environment-induced turbulence, where discs become gravitationally unstable and spiral-arm-dominated within the first $\sim$\,10\,kyr after protostar formation \citep{xu_kunz_2021a_singlecore, xu_kunz_2021b_singlecore}.
We attribute our global stability primarily to the strong magnetisation of our discs in the ideal MHD limit (plasma-$\beta \sim 0.1$, Appendix\,\ref{appendix:complementary_analysis}, Fig.\,\ref{fig:s180cm5_plasma_beta}), which suppresses the build-up of the high surface densities required for fragmentation.

Qualitatively, including non-ideal MHD would likely shift our models toward a more fragmentation-prone regime.
\citet[][their Fig.\,2]{wurster_bate_2019} find that, under non-ideal MHD conditions, fragmentation and spiral structure occur preferentially for rapid rotation ($\beta_{\rm rot}\gtrsim0.035$) and weak magnetic fields ($\mu\gtrsim5$), while stronger fields inhibit fragmentation. Among our nine systems, magnetisation is relatively strong with $\mu \lesssim 3$, but the mean rotation parameter is $\beta_{\rm rot} = 0.063$, which would place our discs in a marginally fragmenting regime, according to the classification of \citet{wurster_bate_2019}.

In general, including radiative transfer in disc formation simulations raises disc temperatures and therefore increases thermal support, which reduces disc fragmentation \citep{commercon_2010_radiative_transfer, lebreuilly_2024a, cusack_2026}. 

Whether or not a disc fragments and in which sizes depends then on the balance between radiative transfer and non-ideal MHD effects, and cannot be determined without dedicated non-ideal radiation-MHD simulations in a cloud-fed setting.


\section{Implications for planet formation}\label{sec:discussion_implications_for_planet_formation}

Our simulations show highly dynamic young discs, characterised by high gas accretion rates onto the outer parts of the disc, and through the disc, leading to continuous replacement of the full disc mass reservoir on short timescales (Fig.\,\ref{fig:disk_masstransport_1D_vertical}).
Such a dynamic mode of accretion challenges models of early dust transport in the solar protoplanetary disc starting from distinct spatially-separated isotopic reservoirs \citep{nanne_2019_viscous_spreading,morbidelli_2022, colmenares_2024}. 
Both the turbulent nature of accretion of filaments onto the disc and the high diffusivity in the disc prevent spatial separation, and even if established, it would be erased through continuous replenishment \citep{ahmad_2026_radtrans_ambi_newNumerical_study}. Possibly, however, at later times not explored here ($t\gtrsim 10^5$\,yr), when infall and accretion rates diminish, disc replenishment timescales exceed disc lifetimes, initiating a more quiescent disc evolution. Late streamer accretion could then inject material with distinct isotopic compositions in the disc \citep{van_Kooten_2024}, possibly paired with an epoch of streamer-induced planetesimal formation \citep{zhao_2025_streamer_pressure_bump}, provided the streamers do not induce strong enough turbulence to suppress planetesimal formation by the streaming instability 
\citep{carrera_2015_SI_in_turbulence, lim_2024_SI_in_turbulence}. 

In general, these young discs in our simulations are not conducive to planet formation. 
Dust coagulation is hindered by the high degree of turbulence, leading to efficient fragmentation down to sub-mm sized particles \citep{vorobyov_2023}. 
Such a high-level of background turbulence may also prevent efficient planetesimal formation \citep{carrera_2025_turbulentclass0_streamers_planetformation}, although previous work has shown the streaming instability to operate in magnetised turbulence \citep{johansen_2007, yang_2018, Xu_Bai_2022, eriksson_2026_SI_under_turbulence}. 
Further growth could be hindered by turbulence-induced planetesimal stirring \citep{nelson_2010, yang_2012} and reduced pebble accretion rates in the stirred-up pebble layer \citep{lambrechts_2012, ormel_2018}.
Therefore, planet formation, at least in the outer parts of protoplanetary discs, requires more quiescent conditions likely found at later, less infall-dominated times. 
Indeed, such lower levels of turbulence are inferred in the outer parts of evolved discs \citep{villenave_2025}.


\section{Conclusion}\label{sec:conclusion}

We present a selected sample of nine star-and-disc systems that formed during two Myr of evolution within a  ($4$\,pc)$^3$
dynamical star-forming region
with densities, velocities, and magnetic field properties representative of nearby star-forming clouds (Sec.\,\ref{sec:resIC}).
Using zoom-in simulations, we simulate their first $10^5$\,yr of disc evolution with a resolution down to $0.8$\,au, employing ideal MHD.
Our main findings are:
\begin{itemize}
  \item The cores in which stars and discs form are turbulent and characterised by a high degree of velocity and magnetic field dispersion (Sec.\,\ref{sec:resutls_the_cores}).
  This results in intermittent filamentary accretion towards the star (Sec.\,\ref{sec:results_stellar_accretion}-\ref{sec:disc_formation_evolution}). 
  The most dense and massive of these filaments can be identified as streamers that can lead to epochs of highly anisotropic mass flow onto the disc. We identify several streamers with lengths of $\sim$\,3000\,au and masses of around 0.2\,M$_\odot$ that drive inwards mass flows at a rate of the order of 10$^{-5}$\,M$_\odot$\,yr$^{-1}$ (Sec.\,\ref{sec:results_streamers}). 
  %
  \item During the disc formation phase ($t\lesssim50$\,kyr), the polar regions near the star display outflows reaching velocities of order 10\,km\,s$^{-1}$, consistent with typical rotational velocities in Keplerian discs at the resolution limit of 0.8 au. These outflows create polar cavities and largely prevent the accretion of low-angular momentum material along the polar axis (Sec.\,\ref{sec:results_flow_and_transport}).
  %
  \item  After the disc formation phase ($t\gtrsim50$\,kyr), the surface layers above and below the disc midplane are dominated by inwards flowing material, supplying $\sim$ $10^{-5}$\,M$_\odot$\,yr$^{-1}$ of gas to the star (Sec.\,\ref{sec:results_flow_and_transport}), replenishing the outer disc on timescales of $\tau_{\rm rep}^{\rm outer}\approx10$\,kyr (Sec.\,\ref{sec:results_mass_transport}). Connected to the high mass-flux surface-layer is a toroidal magnetic field morphology, displaying a field reversal across both disc surfaces and a current sheet within the disc (Sec.\,\ref{sec:magnetic_fields}).
  %
  \item Disc masses reach $M_{\rm disc}\sim10^{-2}$\,M$_\odot$, during the 10$^{5}$\,yr integration period (Sec.\,\ref{sec:all_disc_masses}).
  Growth in disc mass predominantly occurs through increasing the disc size towards $\sim$$100$\,au scales (Sec.\,\ref{sec:all_disc_sizes}), while maintaining a nearly constant surface density profile similar to the MMSN (Sec.\,\ref{sec:results_surfac_density}). 
  Systems are typically gravitationally stable with a disc-to-star mass ratio below 10\%.
  \item When massive destroyer-class streamers interact with the systems, they, on average, truncate the discs, decreasing in size by 65\,\%, while losing 40\,\% of their mass (Sec.\,\ref{sec:results_disk_diversity}).
  Eight out of nine systems exhibit episodic accretion, where stellar accretion rates go from $\dot M_{\rm star} = 10^{-5}$\,M$_\odot$ yr$^{-1}$ to 10$^{-4}$\,M$_\odot$\,yr$^{-1}$, on timescales of about $100$\,yr. In some cases, these transient accretion bursts can be directly linked to the onset of streamer infall (Sec.\,\ref{sec:results_stellar_accretion}).
\end{itemize}

With this work, we show that even relatively isolated star-disc systems inherit different core properties from the turbulent molecular cloud and therefore follow diverse evolutionary paths.
Common to all systems, however, is that young discs are continuously fed cloud material and that discs are repeatedly replenished by accretion flows predominantly along the disc surface.
Additionally, streamers provide an important interplay between the environment and the disc -- with massive destroyer-class streamers truncating discs and re-arranging their angular momentum budget. 
Future work is needed to explore the role of disc and envelope ionisation driving non-ideal MHD effects, the role of outflows launched close to the inner disc edge, radiative feedback, and the initial stages of dust transport and growth inside young discs.

\begin{acknowledgements}
    M.\,L.\,acknowledges funding by the European Research Council (ERC Starting Grant 101041466-EXODOSS).
    The work by M.\,K.\, is funded by the Independent Research Fund Denmark (DFF Sapere Aude Grant: 5251-00016B) and a Carlsberg Reintegration Fellowship (CF22-1014).
    A.\,J. acknowledges funding from the Carlsberg Foundation (Semper Ardens: Advance grant
    FIRSTATMO).
    T.\,H.\,acknowledges funding from the Independent Research
    Fund Denmark through grant No. DFF 10.46540/4283-00305B.
    The Tycho supercomputer hosted at the SCIENCE HPC center at the University of Copenhagen was used for carrying out the simulations and the analysis.
\end{acknowledgements}

\vspace{3cm}

\FloatBarrier 
\bibliographystyle{aa}
\bibliography{references}

\begin{appendix} 

\section{Complementary methods}\label{sec:appendix_methods}

\subsection{MHD equations}\label{appendix:MHD_equations}
The ideal MHD equations are given by the conservation laws for mass,
\begin{align}
    \frac{\partial \rho}{\partial t} 
    +\nabla \cdot (\rho \Vec{v})
    = 0,
    \label{eq:continuity_equation}
\end{align}
and momentum

\begin{align}
    \frac{\partial \rho\Vec{v}}{\partial t}
     + 
    \nabla \cdot \left[\rho\Vec{v}\otimes\Vec{v}
    +
    P_\text{tot} \mathbbm{1}
    +\frac{\Vec{B}\otimes\Vec{B}}{4\pi}
    \right]= -\rho\nabla\Phi 
    \label{eq:momentum_conservation}
\end{align}
Here, $\rho$, $\Vec{v}$ and $\Vec{B}$ are, respectively, the gas density, the velocity field vector and the magnetic field vector. The gravitational acceleration is expressed by $\nabla\Phi$.
The total pressure is given by 
\begin{align}
     P_\text{tot} = P+\frac{B^2}{8\pi}\,.
    \label{eq:total_pressure}
\end{align}
Energy conservation is expressed by
\begin{align}
\frac{\partial E_\text{tot}}{\partial t}
    +\nabla \cdot \left[(E_\text{tot}+P_\text{tot})\mathbbm{1}\Vec{v} - (\Vec{B}\cdot\Vec{v})\frac{\Vec{B}}{4\pi}\right] =-\rho(\Vec{v}\cdot\nabla\Phi)\,,
    \label{eq:energy_equation}
\end{align}
with the total energy being
\begin{align}
    E_\text{tot} = \rho\epsilon + \frac{1}{2}\rho v^2+\frac{B^2}{8\pi}\,.
    \label{eq:total_energy}
\end{align}
It consists of the internal energy density $\rho\epsilon$, the bulk kinetic energy density $1/2\rho v^2$, and the magnetic energy density $B^2/8\pi$ (in cgs units). In this work, we do not solve the energy equation, Eq.\,(\ref{eq:energy_equation}). The thermal evolution of the gas is instead prescribed through a piecewise barotropic equation of state (Sec.\,\ref{appendix:Eos}), used to approximate the thermodynamics without explicitly evolving the internal energy.
Lastly, these MHD equations are solved with a MUSCL/HLLD method similar to the one used in \textsc{ramses} \citep{Fromang2006} (\textsc{dispatch} configuration {\small \verb|SOLVER=ramses/hlld_eos|})

\subsection{Equation of State}\label{appendix:Eos}
We use a piecewise barotropic equation of state,
\begin{equation}
    P = \left\{
\begin{array}{ll}
      c_s^2 \rho  &\quad  \rho\leq\rho_1
      \\
      c_s^2 \rho_1\left(\frac{\rho}{\rho_1}\right)^{1.1}  &\quad  \rho_1 <\rho\leq\rho_2
      \\
      c_s^2 \rho_1\left(\frac{\rho_2}{\rho_1}\right)^{1.1} 
            \left(\frac{\rho}{\rho_2}\right)^{\frac{7}{5}}  &\quad  \rho_2 < \rho\leq\rho_3
      \\
      c_s^2 \rho_1\left(\frac{\rho_2}{\rho_1}\right)^{1.1} 
            \left(\frac{\rho_3}{\rho_2}\right)^{\frac{7}{5}}
            \left(\frac{\rho}{\rho_3}\right)^{1.1}  &\quad  \rho_3 < \rho\leq\rho_4
      \\
      c_s^2 \rho_1\left(\frac{\rho_2}{\rho_1}\right)^{1.1} 
            \left(\frac{\rho_3}{\rho_2}\right)^{\frac{7}{5}}
            \left(\frac{\rho_4}{\rho_3}\right)^{1.1}
            \left(\frac{\rho}{\rho_4}\right)^{\frac{5}{3}}
           &\quad  \rho\ge \rho_4\,,
      \\
\end{array} 
\right.
\label{eq:equation_of_state}
\end{equation}
based on fitting the temperature evolution during core collapse in a radiative hydrodynamic model \citep{masunaga_2000} which has frequently been used in other studies as well \citep{machida_2007_jetlunchpoint,kuruwita_2020, jorgensen_2022_boxused, tuhtan_2023, yang_2025}. Here, the sound speed $c_{s,0}$ is set to be $c_{s,0} = 1.8 \times 10^4$\,cm\,s$^{-1}$, assuming $T=10$\,K and mean molecular weight of $\mu=2.34\,m_{\rm H}$. Here,  $m_{\rm H}$ is the mass of the hydrogen atom. The transition densities are 
$\rho_1=2.50 \times10^{-16}$, 
$\rho_2=3.84 \times10^{-13}$,
$\rho_3=3.84 \times10^{-8}$, 
and $\rho_4=3.84 \times10^{-3}$\,g\,cm$^{-3}$.
A density of $\rho_1=2.50 \times10^{-16}$\,g\,cm$^{-3}$, equal to a number density of $n_1=10^{7}$\,cm$^{-3}$, marks the departure from isothermal evolution, as the collapsing gas becomes optically thick to its own thermal (infrared) radiation and can no longer radiate away compressional heating efficiently. The transition at $\rho_2=3.84 \times10^{-13}$\,g\,cm$^{-3}$ ($n_2\approx 10^{11}$\,cm$^{-3}$) marks the onset of the adiabatic regime and the formation of the first hydrostatic core \citep{larson_1969}.

\subsection{Sink particle prescription}\label{appendix:sink_cell_prescription}

Sink particles are used in \textsc{dispatch} to provide a sub-grid formulation for gravitationally collapsed objects. A sink particle is created in a cell when a number of criteria, designed to detect gravitational collapse beyond what can be modelled with the given resolution, are fulfilled. Once sink particles are formed, they can acquire mass and momentum through accretion from neighbouring cells. A corresponding amount of mass, momentum, and thermal energy is removed from the cells in each accretion event.

Sink creation is triggered according to the same criteria used in the Copenhagen version of \textsc{ramses} \citep{haugbølle_2018}. We check that (1) the local density exceeds the threshold corresponding to 2 cells per Jeans length, (2) the velocity field is converging, (3) the gravitational potential is at a local minimum, and (4) the sink particle is created at a distance of at least 200 cells from any other sink particle. The distance criterion could be relaxed down to 8 cells, but we set it to a high value, given the isolated nature of the sinks selected for zoom-in.

Sink particles accrete from neighbouring cells inside a radius $r_\text{acc}$, in our case set to 8 cells. The overall time-scale is set by a combination of rotational and free-fall timescales, by summing the corresponding rates. A base fraction of material that is to be accreted, is determined as
\begin{equation}
    \epsilon = (\Omega_\mathrm{disc} + \Omega_\mathrm{free-fall}) \Delta t\,,
\end{equation}
where $\Delta t$ is the current timestep. The frequencies are related to disc-mediated and free-fall accretion. The disc accretion rate is a fraction of the Keplerian rotation rate measured at $r_\text{acc}$
\begin{gather}
    \Omega_\mathrm{disc} = \sqrt{\frac{GM_*}{{r_\text{acc}}^3}} \cdot \chi_\mathrm{disc}\,,
    \label{eq:omega_disk}
\end{gather}
where $M_*$ is the mass of the sink particle and $\chi_\mathrm{disc}$ is the disc accretion efficiency. The free-fall accretion rate is given by 
\begin{gather}
    \Omega_\mathrm{Free-fall} = \frac{\chi_\mathrm{ff}}{t_{ff}} = \sqrt{\frac{32G\rho_\text{max}}{3\pi}}\cdot\chi_\mathrm{ff}\,
    \label{eq:omega_free-fall}
\end{gather}
where $\rho_\text{max}$ is the maximum density within the accretion sphere and $\chi_\mathrm{ff}$ is the free-fall accretion efficiency. The fraction of material accreted from a single cell, $\Delta_i$, is related to the base fraction for accretion towards the sink particle, but modulated with five different factors. It is (1) limited by a maximum fraction $\epsilon_\mathrm{max}$, (2) gradually decreased towards the edge of the accretion radius with a quartic function of the distance to the sink particle, $r_i$, (3) decreased according to the density in the cell, normalised by the average density inside the accretion radius, $\langle\rho\rangle$, (4) decreased if cells have velocities pointing away from the sink. Furthermore, (5) we include a safety valve that always accretes material such that the density in any given cell inside the accretion radius does not go above a maximum density $\rho_\textrm{acc}$, in our case set to 100 times above the limit for creating new sink particles. The final accretion rate from any cell inside the accretion radius may thus be written as

\begin{align}  \nonumber
    \Delta & = \epsilon  \frac{\epsilon_\mathrm{max}}{\epsilon_\mathrm{max} + \epsilon}
     \left(1-\frac{r_i^2}{{r_\text{acc}}^2}\right)^2  \left(1 - \exp\left(- \frac{\rho_i}{\langle\rho\rangle}\right)\right) \\ \label{eq:sink_accretion}
           & \quad  \cdot \quad \left\{
    \begin{array}{ll}
      0 & 1 < \tan\theta 
      \\
      1 - \left(\tan\theta\right)^2  & 0 < \tan\theta < 1 
      \\
      1 &  0 > \tan\theta,
      \\
    \end{array}    
    \right\} 
\end{align}

\begin{align}
    \Delta_i  = \left\{ 
     \begin{array}{ll}
      \Delta & \Delta > \frac{\rho_c \chi_\rho}{\rho_i} \\
      1 - \frac{\rho_c \chi_\rho}{\rho_i} & \Delta < \frac{\rho_c \chi_\rho}{\rho_i} \\
    \end{array}
    \right. \,,
\end{align}
where $\tan\theta=v_r/|v_{\rm ang}|$ and $v_{\rm ang}$ is the non-radial part of the velocity (relative to the sink), hence perpendicular to $v_r$. This accretion recipe is similar, while still slightly different from the accretion prescription in \cite{haugbølle_2018}.
In each timestep, the sink particle properties are then updated as
\begin{align}
\Delta M_\textrm{sink} & = \sum_{r_i < r_\textrm{acc}} \Delta_i \rho_i \Delta V_i \\ \nonumber
\vec{\Delta p}_\textrm{sink} & = \sum_{r_i < r_\textrm{acc}} \Delta_i \rho_i \vec{v}_i \Delta V_i\,,
\end{align}
where $\Delta V_i$ is the cell volume of cell $i$. The gas cells inside the accretion radius are updated correspondingly
\begin{align} \nonumber
\Delta \rho_i & = - \Delta_i \rho_i \\ \label{eq:gas_decretion}
\Delta \rho_i \vec{v}_i & = - \Delta_i \rho_i \vec{v}_i \\  \nonumber
\Delta e_i  & = - \Delta_i e_i\,,
\end{align}
where $\Delta \rho_i$, $\Delta \rho_i \vec{v}_i$, $\Delta e_i$ are the changes in density, momentum, and thermal energy in cell $i$. Magnetic fields are not accreted.

In our models, we aim to use sink parameters that result in a relatively undisturbed density profile towards the sink. Some level of impact is unavoidable and apparent, such as the plateau of constant surface density towards the accretion radius, as seen in panel b of Figure\,\ref{fig:s180cm5_gendisk}. We set $\chi_{\rm disc}=10^{-3}$, corresponding to inwards spiralling motion at a rate of 0.1\% per radian of rotation at the accretion radius, and $\chi_{\rm ff}=1$, corresponding to free-fall accretion when there is no disc.
The accretion radius is set to 8 cells at the highest refinement level $r_{\rm acc}=8\,ds_{\rm min}$ and the maximum accretion fraction is set to $\epsilon_{\rm max}=80\,\%$, which appears to be a fair approximation of the resolution limitation on the outflow rate (cf.\,below). The maximum allowed density in the accretion region is $\rho_c \chi_\rho$, with $\chi_\rho=100$. This ``safety valve'' density level, which would indeed correspond to a Jeans' resolution of only 0.2 cells, is never reached in our runs.

\subsection{Numerical Resolution}\label{appendix:numerical_resolution}
\begin{figure}
    \centering
    \resizebox{0.8\hsize}{!}{\includegraphics{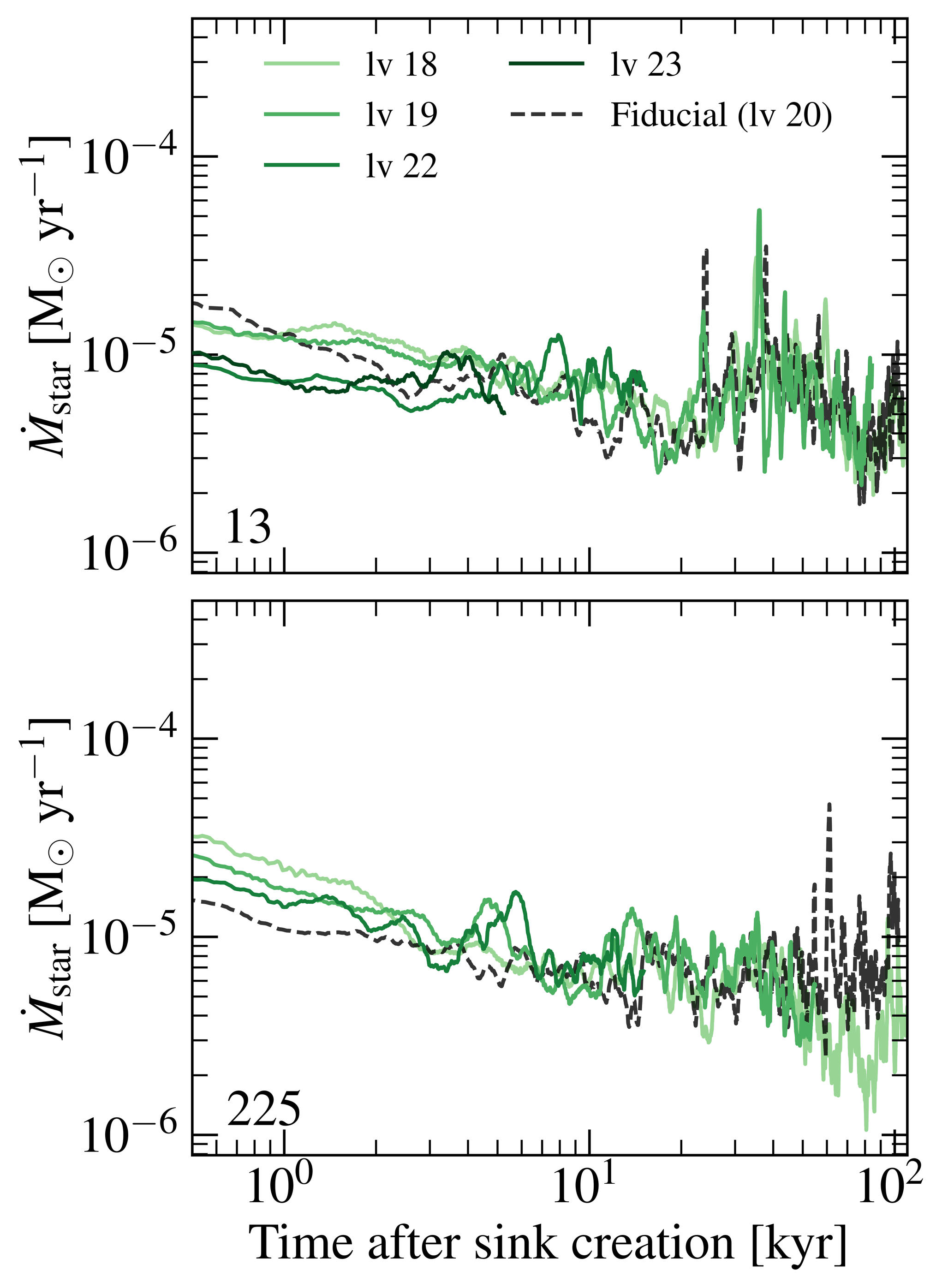}}
    \caption{Accretion rates for systems \texttt{13} and \texttt{225} with varying maximum level parameter. The refinement parameter was tested with 5 settings for system \texttt{13} and 4 for \texttt{225}.}
    \label{fig:refinement_accretion}
\end{figure}

This project was carried out with a maximum resolution of 0.8 au at level 20; this was chosen as a compromise between a reasonable run time and highest achievable resolution (all runs were carried out on single 128 core EPYC nodes using 32 or 64 OpenMP threads). 

In Figure\,\ref{fig:refinement_accretion}, a small study of different refinement settings is shown. For system \texttt{13}, the resolution was increased to levels 22 and 23, i.e. with the smallest cells at 0.2 au and 0.1 au, respectively. For system \texttt{225}, level 22 was the highest resolution. In both panels, the fiducial run at level 20 is marked with the black, dashed line. In both panels, levels 18 and 19 are showing, having the smallest cell sizes at 3.2 au and 1.6 au, respectively. 

As can be appreciated in Figure\,\ref{fig:refinement_accretion}, accretion rates decrease with increasing levels of refinement (smaller cell size). This is mainly because the total mass flux in the outflow increases with resolution, and only converges when the launch point of the jet is marginally resolved. The launch point for Solar mass stars is at $\sim$ 0.01 au \citep{machida_2007_jetlunchpoint}. Furthermore, at higher resolution, higher outflow speeds are observed \citep{tu_2026_highvelocityjets}. This is because the extracted angular momentum available in an outflow is proportional to the Keplerian angular momentum in the disc, with some lever arm \citep{spruit_1996}. From our convergence studies, and other zoom-in simulations done in the past, we estimate that with a 0.8 au resolution, the stellar mass is overestimated by up to 30\,\%.

\subsection{Fitting disc radii}
\label{app:radfit}
In our analysis, we combine two physical criteria to determine the disc radius that are related to rotational support by the central object and structural stability. We find that the combination of these two physically motivated constraints creates a robust measure that aligns very well with the disc radius that can be estimated by eye when considering discs in a diverse set of evolutionary states.

To determine a disc size, the disc is segmented into radial bins. For our analysis, we use 500 logarithmically spaced bins from 5\,au to 500\,au. The first condition is that the force balance on individual gas elements is dominated by the gravitational pull of the central object, formulated in the condition
\begin{align}
    v_\phi > 0.8\,v_{\rm kep}\,,
\end{align}
where $v_{\rm kep}$ is the Keplerian velocity and $v_\phi$ is the mass-weighted azimuthal velocity measured in a radial bin and within three scale heights from the midplane. The second condition is that the Rayleigh stability criterion for differentially rotating fluids \citep{rayleigh_1917} is fulfilled
\begin{align}
    \Phi = \frac{1}{r^3}\frac{d}{dr}(r^2\varOmega)^2\,>0\,.
\end{align}
In practice, applying these conditions requires a certain amount of tuning to be robust. For example, large features, such as spiral arms, in the disc can trigger the disc edge criterion and to avoid this, we filter the measured quantities. Please see the \textsc{radvis} analysis pipeline for details: \url{https://github.com/CGHolm/RaDvisPython}.

By incorporating both criteria, we ensure that the disc edge is defined by a physical transition from a supported rotating flow to an unsupported flow, while maintaining a near-Keplerian velocity. We propagate the standard deviation of $v_\phi$ and $\Phi$ in each radial bin using the inverse of the gradient at the disc size radius to provide an uncertainty for the disc size. E.g. $\sigma_{r, \Phi}=\left|\frac{\partial \Phi}{\partial r}\right|^{-1}\sigma_\Phi$. We combine the two disc size measurements using a weighted mean
\begin{equation}
r_\textrm{disc\,edge} = \frac{\sigma_{r, \Phi}^{-2} \,r_{\Phi} + \sigma_{r, v_\phi}^{-2}\, r_{v_\phi}}{\sigma_{r, \Phi}^{-2} + \sigma_{r, v_\phi}^{-2}} 
\end{equation}
to calculate the final disc size.

\subsection{Disc fitting procedure}\label{sec:appendix_disc_fitting_procedure}
When fitting the surface density and scale height, the spin axis is initially calculated for a 50\,au sphere. Using the spin axis as the new $z'$-axis, cylindrical basis vectors are assigned to all cells. Within this new coordinate system, we calculate the angular momentum only considering the disc midplane, i.e. ($r<150$\,au $\bigwedge$ $|z'|<15$ au) $\bigvee$ $|z'|/r<0.3$. The last two steps are then repeated until the spin axis converges $\Delta\theta<5^\circ$. 
We characterise the 1D radial structure of the disc, by assuming hydrostatic balance
\begin{align}
    \rho(r,z)=\frac{\Sigma(r)}{\sqrt{2\pi}H}  \exp{\left(-\frac{z^2}{2H^2}\right)}.
    \label{eq:vertical_structure}
\end{align}
across a midplane ($z=0$) perpendicular to the spin axis of the disc. 
Here, $\Sigma$ and $H$ are, respectively, the best-fit gas surface density and the gas scale height for the density profile $\rho(r,z)$ of the disc.

\section{Complementary analysis}\label{appendix:complementary_analysis}
\subsection{Virial Theorem}\label{appendix:virial_theorem}
\begin{figure}
    \centering
    \resizebox{0.95\hsize}{!}{\includegraphics{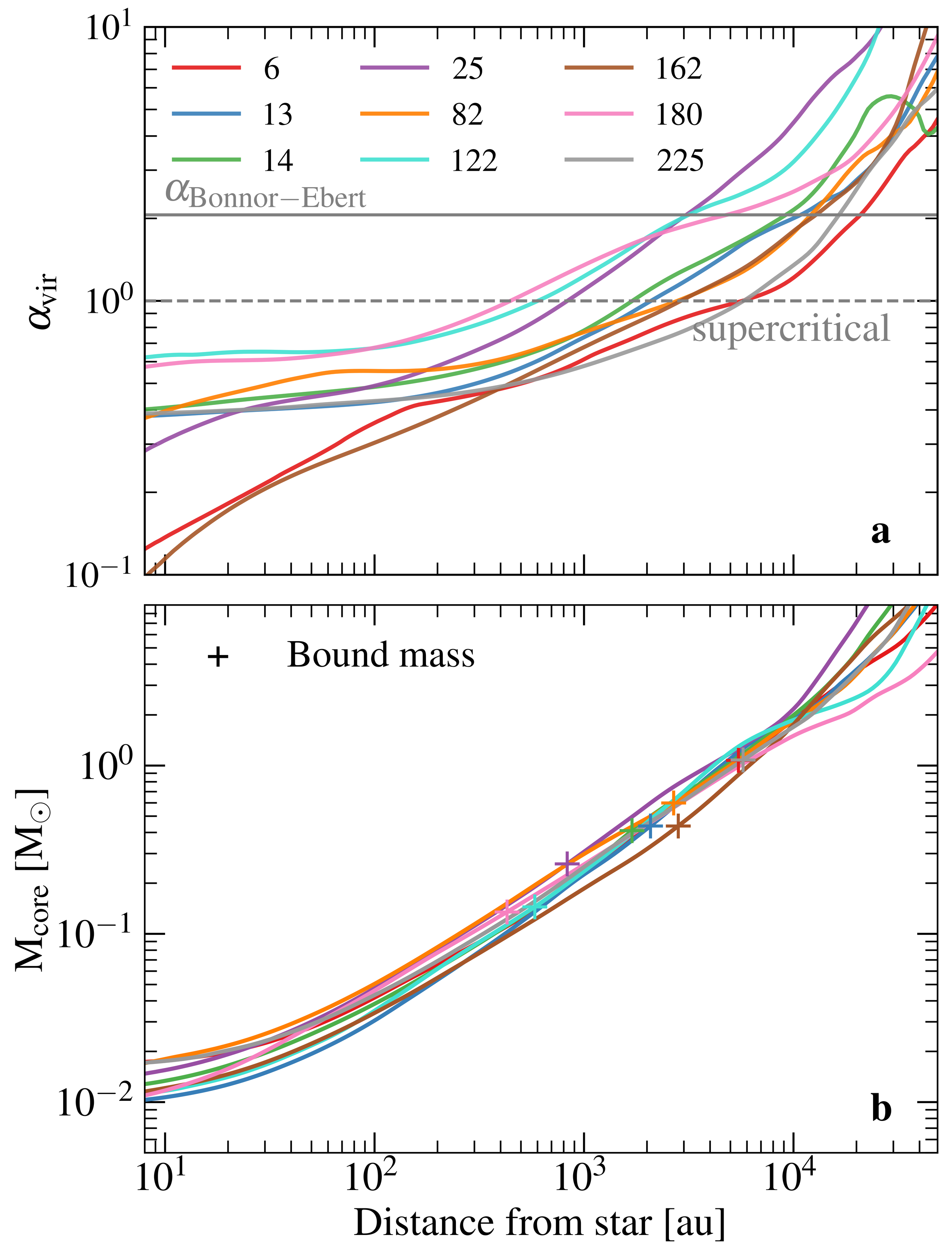}}
    \caption{ Panel a: 
    Virial number as a function of radius
    for the nine cores at $T_0$ (see row 2 in Table\,\ref{tab:core_overview}). Panel b: Cumulative mass as a function of distance to the star. Crosses indicate the initial bound mass in accordance with $\alpha_{\rm vir}$ in panel a. 
    The plot shows no bound material for our selected cores beyond $\sim10^4$\,au.}
    \label{fig:non-alpha_vir}
\end{figure}

The selected cores are all supercritical out to $10^3$\,au
with no cores having bound material beyond $10^4$\,au 
(Fig.\,\ref{fig:non-alpha_vir}). 
We use the virial number to determine the initial bound material of the cores, and it is defined as, 
\begin{equation}
\alpha_{\rm vir}=  \frac{2(E_{\rm kin}+E_{\rm therm})}{E_{\rm grav}}.
\end{equation}
The virial number is the ratio of bulk kinetic and thermal energy to gravitational energy. With dominating gravitational energy ($\alpha_{\rm vir}<1$), the cores are supercritical and prone to collapse \citep{bertoldi_1992_virialtheorem}.  
The supercritical core masses 
lie between 0.1 - 1.1\,M$_\odot$ (Fig.\,\ref{fig:non-alpha_vir}, panel b). 
We also show the virial number, $\alpha_{\rm Bonnor-Ebert}$, with characteristic values for a Bonnor-Ebert sphere presented in \citet{kauffmann_2013_alphavir}. Only systems \texttt{6} and \texttt{225} have $\alpha_{\rm vir}=\alpha_{\rm Bonnor-Ebert}$ at $\sim10^4$\,au.

\begin{table*}
    \centering
    \caption{\label{tab:mass_to_flux_overview} 
    Mass-to-flux ratios for the nine different cores. For each system, we calculate the flux for three planes aligned with the spin axis and the volume-averaged magnetic field direction. Column 2 gives the planes $yz$, $zx$ and $xy$. The angle between $ \Vec{B}$ and $ \Vec{L}$ is found in row 8 and the bottom row. All values are computed at $R=10^4$\,au and $R=10^3$\,au.
    }
    \setlength{\tabcolsep}{4pt} %
    \begin{tabular}{llcccccccccr}
    \hline\hline
      \multirow{2}{*}{$R=10^4$\,au\hspace{-4cm}} & 
      \multirow{2}{*}{} & 
      \multirow{2}{*}{\texttt{6}} & 
      \multirow{2}{*}{\texttt{13}} & 
      \multirow{2}{*}{\texttt{14}} &
      \multirow{2}{*}{\texttt{25}} &
      \multirow{2}{*}{\texttt{82}} &
      \multirow{2}{*}{\texttt{122}}&
      \multirow{2}{*}{\texttt{162}}&
      \multirow{2}{*}{\texttt{180}}& 
      \multirow{2}{*}{\texttt{225}}&
      \multirow{2}{*}{1}
      \\
      &
      &
      &
      &
      \\ \hline           
        \multirow{3}{*}{$z= \Vec{L}$} & $\mu(yz_\perp)$ & 8.24 & 3.01 & 3.38  &  2.29 &  7.86  & 6.49 & 2.30 & 3.29 & 3.69 & 2\\
         &                             $\mu(zx_\perp)$ & 14.0 & 2.76 &  148  &  1.87 &  12.5  & 2.30 & 3.42 & 8.76 & 5.15 & 3\\
         &                             $\mu(xy_\perp)$ & 3.11 & 31.1 &  13.2 &  2.21 &  1.61  & 6.77 & 8.65 & 7.75 & 1.83 & 4\\  \hline
        \multirow{3}{*}{$z= \Vec{B}$} & $\mu(yz_\perp)$ & 151  & 226  &  214$\times10^1$ &  114$\times10^1$ & 22.4 & 147 & 467 & 101 & 28.9 & 5\\
        &                              $\mu(zx_\perp)$ & 20.2 & 45.3 &  102  &  30.3 &  197$\times10^1$ & 61.7 & 606 & 143 & 47.8  & 6\\
        &                              $\mu(xy_\perp)$ & 2.57  & 2.19 &  3.12  &  1.28 & 1.51 & 2.01 & 1.88 & 2.67 & 1.49  & 7\\  \hline
        $\theta( \Vec{B}, \Vec{L})$&      [deg]          & 162   & 86.2 &  76.3  &  124 &  15.0 & 70.5 & 77.3 & 65.5 & 147 & 8 \\  
        \hline
     \multirow{2}{*}{$R=10^3$\,au} & & & & & & & & & & &  \multirow{2}{*}{9}\\
       &  \\ \hline
         \multirow{3}{*}{$z= \Vec{L}$} & $\mu(yz_\perp)$ & 17.8 & 17.2 & 7.45  & 18.8 &  9.74  & 29.6 & 20.4 & 16.4 & 4.85& 10 \\
         &                             $\mu(zx_\perp)$ & 4.06 & 6.20 &  25.1  &  17.2&  3.00  & 5.55 & 441 & 53.1 & 5.57&11 \\
         &                             $\mu(xy_\perp)$ & 80.5 & 5.56 &  5.28 &  2.73 &  41.8  & 8.95 & 166 & 4.99 & 35.5 &12\\\hline
        \multirow{3}{*}{$z= \Vec{B}$} & $\mu(yz_\perp)$ & 753  & 63.2  &  22.6 & 736  & 103    & 166  & 118 & 391  & 110&13\\
        &                              $\mu(zx_\perp)$ & 164  & 89.2 &  66.6  &  125 & 852    & 71.7 & 455 & 111 & 88.3  &14\\
        &                              $\mu(xy_\perp)$ & 3.44  & 4.11 &  3.65  &  2.67 & 2.61 & 4.34 & 4.27 & 4.76 & 3.50  &15\\  \hline
        $\theta( \Vec{B}, \Vec{L})$&      [deg]          & 82.4  & 45.5 &  138  &  167 &  85.7 & 51.7 & 155  & 19.6 & 93.4  &16\\  
    \hline\hline
  \end{tabular}
\end{table*}

\subsection{Mass-to-flux ratio}\label{sec:appendix_masstoflux}
The mass-to-flux ratio is highly dependent on geometry. We present 12 mass-to-flux measurements for each system at sink creation time ($t=0$\,kyr), in Table\,\ref{tab:mass_to_flux_overview}. We measure the flux through each surface perpendicular to the three basis vectors with respect to the spin axis and the volume-averaged mean magnetic field direction. This is done for a sphere with radius $R=10^4$\,au and $R=10^3$\,au. In Table\,\ref{tab:mass_to_flux_overview} we also report the angle between the mean B-field direction and the spin axis (row 8 and 16).

We find that, on the largest scale ($10^4$\,au), only three systems (\texttt{6}, \texttt{82}, and \texttt{225}) have alignment between the spin axis and the highest measured flux, yielding the lowest mass-to-flux ratio through the $xy$-plane ($\Vec{z}\equiv\Vec{L}$). Naturally, the lowest mass-to-flux ratio is obtained along the mean field direction (row 5-7, Table\,\ref{tab:mass_to_flux_overview}), emphasising the importance of geometry when calculating the mass-to-flux ratio. The importance of calculating the flux along the magnetic field is further supported for the inner core ($R=10^3$\,au), seen in Table\,\ref{tab:mass_to_flux_overview} row 10-15. At 10$^3$ au the average mass-to-flux ratio across the cores is 3.7\,$\pm$\,0.7. This trend reflects increasing magnetic support in the outer envelope and highlights that the choice of scale and core definition impacts the inferred mass-to-flux ratio. 
Comparing our values to those reported in \cite{kuffmeier_2017} again reveals lower mass-to-flux ratios, suggesting we are investigating more magnetically supported cores, and once again emphasising the importance of geometry when measuring flux. We note the inclination between spin axis, $\Vec{L}$, and volume-average B-field direction is categorised as large misalignments, in relation to old core collapse models \citep{joos_2012_magnetic_breakning_catastrophe, Li_2014_MHD_review}.

\subsection{Plasma-$\beta$}
\begin{figure}
    \centering
    \resizebox{0.95\hsize}{!}{\includegraphics{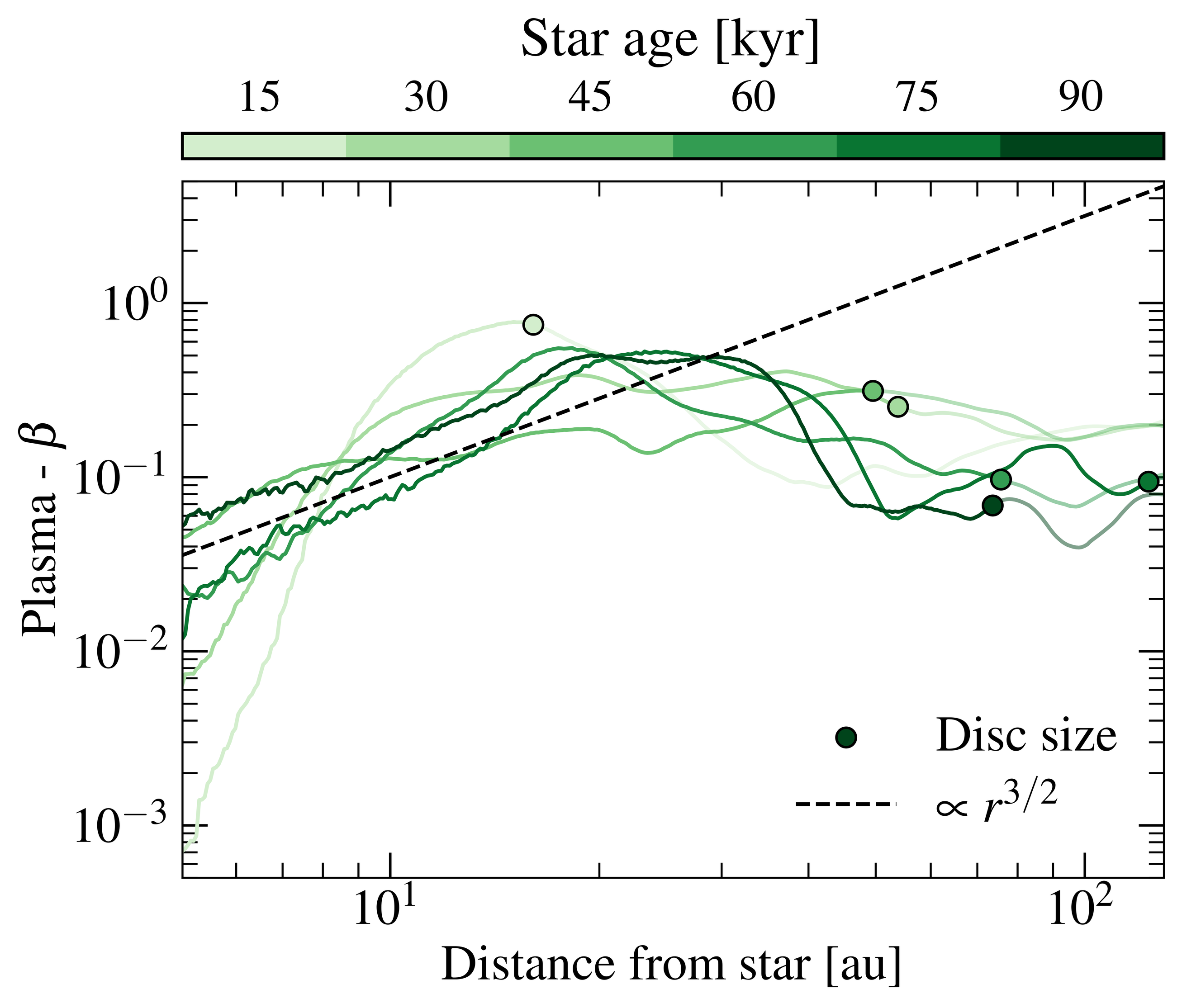}}
    \caption{Plasma-$\beta$ as function of radial distance, for system \texttt{180} at different times. The values are calculated in 500 logarithmically-spaced radial cylindrical bins. For each bin, we measure the non-weighted $\beta$ average within $\pm1 H$, where we obtain the highest resolution.} 
    \label{fig:s180cm5_plasma_beta}
\end{figure}

We obtain values canonical for ideal MHD simulations of plasma-$\beta<1$ \citep{kuffmeier_2024_review}. We plot the plasma-$\beta$ values in Figure\,\ref{fig:s180cm5_plasma_beta}. The plot shows how the inner disc shows lower plasma-$\beta$ values. This is highly coupled with the efficient depletion of material in the inner radii (panel c, Fig.\;\ref{fig:s180cm5_gendisk}).

\end{appendix}

\end{document}